 \documentclass[twocolumn,notitlepage,nofootinbib,pra,superscriptaddress]{revtex4-1}
\usepackage{amsmath}
\usepackage{amsfonts}
\usepackage{amssymb}
\usepackage{graphicx}
\usepackage{sidecap}
\newcommand{\sign}{\text{sign}}
\usepackage{braket}

\usepackage{blindtext}
\usepackage{subfigure}
\usepackage{lipsum}
\usepackage{fullpage}
\newcommand*\diff{\mathop{}\!\mathrm{d}}
\usepackage{appendix}
\usepackage{subfigure}
\usepackage{color}
\usepackage[dvipsnames]{xcolor}
\usepackage{physics}

 \usepackage{amsthm}

\newtheorem{theorem}{Theorem}
\newtheorem{rules}{Rule}

\begin{document}
\title{Diagrammatic approach for analytical non-Markovian time-evolution: Fermi's two atom problem and causality in waveguide quantum electrodynamics}
\author{Fatih Dinc}
\email{fdinc@stanford.edu}
\affiliation{Department of Applied Physics, Stanford University, Stanford, CA 94305, USA}

\begin{abstract}
Non-Markovian time-evolution of quantum systems is a challenging problem, often mitigated by employing numerical methods or making simplifying assumptions. In this work, we address this problem in waveguide QED by developing a diagrammatic approach, which performs fully analytical non-Markovian time evolution of single-photon states. By revisiting Fermi's two atom problem, we tackle the impeding question of whether rotating-wave approximation violates causality in single-photon waveguide QED. Afterward, we introduce and prove the \emph{no upper half-plane poles (no-UHP) theorem}, which connects the poles of scattering parameters to the causality principle. Finally, we visualize the time-delayed coherent quantum feedback mediated by the field, discuss the Markovian limit for microscopically separated qubits where short-distance causality violations occur and the emergence of collective decay rates in this limit. Our diagrammatic approach is the first method to perform exact and analytical non-Markovian time evolution of multi-emitter systems in waveguide QED.
\end{abstract}
\maketitle

\section{Introduction}
Experimental realization of quantum computers is closely linked to the theoretical developments in describing interactions between distant qubits mediated by waveguides \cite{kimble2008quantum}. Aiming to uncover the physics behind such interactions, waveguide quantum electrodynamics (QED) has the potential to aid with the development of quantum technologies in many fronts including but not limited to quantum memory, communication, sensing and routing \cite{corzo2019waveguide,calajo2019exciting,goban2015superradiance,degen2017quantum,lvovsky2009optical,shomroni2014all,sipahigil2016integrated}. On the other hand, the theoretical advancements in waveguide QED are limited by our ability to describe macroscopically separated systems, where time-delayed non-Markovian feedback effects dominate and render many simplifying assumptions irrelevant \cite{grimsmo2015time,nemet2019comparison,guimond2017delayed,pichler2016photonic}.

Theoretical waveguide QED community has developed a multitude of methods to tackle the modeling issues associated with waveguide mediated interactions between qubits. Some examples are real-space approaches for few-photons, input-output formalism for describing asymptotic and steady-states, generalized master equations particularly well-suited for describing many-photons and the SLH formalism with its simple set of rules \cite{zheng2010waveguide,shen2007strongly,bermel2006single,cheng2016single,shen2018exact,dinc2019exact,roulet2016solving,roulet2016solving,caneva2015quantum,rephaeli2013dissipation,rephaeli2013dissipation,fan2010input,shi2015multiphoton,baragiola2012n,shi2009lehmann,shi2011two,combes2018two,brod2016two,gonzalez2013non}. Among those methods, many had been employed to describe interactions between distant qubits; yet, while some perform numerical or semi-analytical calculations \cite{fang2018non,liao2015single,calajo2019exciting,dinc2019exact,lalumiere2013input,chen2011coherent,sato2012strong,albrecht2019subradiant,liao2016dynamical} others perform simplifying assumptions such as N-pole approximation or superradiance condition assumption \cite{sinha2020non,zheng2013persistent}. However, a method for describing exact and fully analytical non-Markovian time-evolution is missing from the literature.

The challenge of performing analytical time-evolution of a given Hamiltonian is not a new concept in the quantum physics literature. Even for introductory examples in quantum mechanics and quantum optics \cite{griffiths2010introduction,grynberg2010introduction}, many elementary problems are solved numerically or with simplifying assumptions. Similarly in \cite{dinc2019exact}, we have shown that the analytical non-Markovian time-evolution in single-photon waveguide QED is one analytically-hard-to-compute integral away from our reach, but available semi-analytically. Like every other difficult problem, we need to shift our perspective and tackle the problem with a solid intuition and a clear motivation. The intuition comes from the locality of interactions between qubits due to point-coupling \cite{trivedi2019point}, whereas the motivation stems from a nearly century-old causality question arisen from Fermi's 1932 paper \cite{fermi1932quantum}, which still needs to be answered for multi-qubit waveguide QED.

A cause cannot have an effect propagating faster than the speed of light, a fundamental principle widely referred to as the \emph{causality}. This principle is at the heart of all quantum field theories, effective or otherwise, yet has been shown to be violated by two central concepts of the quantum electrodynamics (QED) literature: i) the rotating-wave approximation (RWA) performed at the Hamiltonian level  \cite{hegerfeldt1994causality,Compagno_1989,funai2019faster}. ii) the concept of a photon wave function in real-space \cite{sipe1995photon}. On the other hand, causality has emerged \emph{exactly} \cite{fang2018non,xu2015input,dinc2019exact} and approximately \cite{sanchez2018emergent} many times in waveguide QED. 

Despite utilizing a wave-function approach, real-space formalism has not predicted any non-causal spontaneous emission dynamics so far. Similarly, RWA-related non-causality had been cured by another approximation performed by Fermi in his 1932 paper \cite{fermi1932quantum} -- namely the extension of integral bounds to negative energies. Ever since, Fermi's two atom problem has been used as a benchmark to check whether causality in a theory is violated. In 1995, \cite{milonni1995photodetection} has shown that the negative frequency approximation deletes the non-causal terms, which would have been cancelled by the contributions coming from the non-rotating terms. Consequently, many studies in waveguide QED use the negative frequency extension \cite{zheng2013persistent,tsoi2008quantum}. In fact, a recent study shows that the negative-frequency extension leads to the exact and analytical result for the spontaneous emission from an initially excited qubit and is therefore an essential part of the formalism \cite{dinc2019exact}. Another recent study has studied Fermi's two atom problem in circuit QED using Heisenberg picture \cite{sabin2011fermi}, showing that causality is not violated in this setup. However, a rock-solid argument of causality for multi-qubit systems can be made only with analytical time-evolution of general multi-qubit systems.

In this paper, we develop a diagrammatic approach that gives \emph{exact and fully analytical non-Markovian time evolution} in single-photon waveguide QED. We find the detuned (absolute) momentum space diagram rules that can be used to calculate time evolution without explicit computation. By revisiting Fermi's two atom problem equivalent in waveguide QED, we show that non-Markovian waveguide QED is safe from the non-causal behavior discussed in the literature regarding both RWA and photon wave-function. Finally, using causality arguments, we introduce and prove the no upper half-plane poles (no-UHP) theorem, a waveguide QED analogy to previously known causality theorems in quantum theory \cite{gellmann1954use,schutzer1951onthe}. No-UHP theorem can be used to probe whether results of any calculation in waveguide QED respects the causality principle, hence provides a sanity check for future research. 

The main contributions of this paper are to introduce and develop the diagrammatic method, discuss causality in waveguide QED, and prove no-UHP theorem. While other diagrammatic approaches exist in waveguide QED, our method is distinct from them. These diagrams help finding the steady-state solution (S-matrix elements) \cite{see2017diagrammatic,pletyukhov2012scattering}, consider a single-qubit utilizing operational translation of the atomic nonlinear response \cite{roulet2016solving} or provide perturbative solutions in the Heisenberg picture \cite{sabin2011fermi}, our diagrams find exact time-evolution for many-qubit systems utilizing locality principle. To the best of our knowledge, this is the first work considering fully analytical time dynamics of multiple emitters in non-Markovian regime and exact solution of the time-dependent Schr\"odinger equation without any further approximation or assumption.

For the entirety of this work, we encounter two sources of non-Markovianity: i) due to the finite width of single-photon wave-packets \cite{valente2016non,fang2018non} ii) due to finite propagation time between qubits \cite{dinc2019exact,zheng2013persistent,sinha2020non,dinc2019non}. Thus, our consideration of non-Markovianity is uniquely determined by the Schr\"odinger time evolution, system geometry and the initial conditions. As our interests lie in introducing the diagrammatic method for time-evolution and answering the causality question for multi-qubit systems, we do not consider non-radiative decay or any additional noise added to the system, which are natural extensions of the results presented in this paper.

\section{Theory of Diagrams}
In this section, we develop our diagrammatic method starting from the Hamiltonian and the equations of motion. We first show that the equations of motion for the field amplitudes and qubit excitation coefficients (together, we simply call them \emph{amplitudes}) are local owing to the point-coupling nature of interactions. Later, we use the locality principle to divide the scattering processes into unit cells, where a specific \emph{event}, a physical process described by a diagram, can be described by cascading multiple unit cells. We associate a set of rules with each unit cell, which performs the computation of complex integrals automatically. Finally, we use the diagram rules to perform an example amplitude computation for an example diagram and provide a systematic method for finding all diagrams that contribute to the calculation of a specific amplitude.

\subsection{Hamiltonian and equations of motion}
We motivate our approach for a linear chain of $N$ equally-separated (with a distance $L$) identical qubits coupled to a one-dimensional waveguide, although neither assumptions are necessary. The Hamiltonian for such a system can be given as $H = H_0 + H_I$, where
\begin{equation}
    \begin{split}
    H_{0}={}& \Omega \sum_{Q\in\{\rm qubits\}} \sigma_Q^\dag \sigma_Q + i \hbar v_g \int_{-\infty}^\infty \diff x\\
    & \times\left( C_L^\dag(x)\frac{\partial}{\partial x} C_L(x) -C_R^\dag(x)\frac{\partial}{\partial x} C_R(x)  \right)   
    \end{split}
\end{equation}
is the free Hamiltonian. $\sigma_Q^\dag$ is the excitation operator for the qubit $Q$, $v_g$ is the group velocity (for simplicity $v_g=\hbar=1$), $C_{R/L}^\dag(x)$ are creation operators for right/left moving photons. Similarly,
\begin{align}
    \begin{split}
    H_{I}={}&\sqrt{\gamma_0/2} \sum_{Q\in\{\rm qubits\}} \int_{-\infty}^\infty \diff x \delta(x-QL)\\
    &\times\left( (C_R^\dag(x)+C_L^\dag(x)) \sigma_Q + \text{H.c.} \right)\,,
    \end{split}
\end{align}
is the point-coupling interaction Hamiltonian with $\gamma_0=2J_0$ being the single emitter decay rate\footnote{Throughout this paper, we sometimes use $J_0$ over $\gamma_0$ when it is algebraically convenient.}. There are several assumptions of the real space Hamiltonian \cite{dinc2019exact}: The rotating-wave approximation is employed when obtaining $H_I$, we assume that we are interested in energy levels $E_k = \Omega \pm O(\gamma_0)$ with $\gamma_0 \ll \Omega$. 

In \cite{dinc2019exact}, we considered the time evolution of single-excitation states by re-constructing the time-evolution operator using the energy-eigenstates. It has lead us to finding exact and analytical time-evolution under the Markovian approximation and semi-analytical time-evolution for non-Markovian systems. In this paper, to find analytical time-evolution for general non-Markovian systems, we consider a general single-excitation state
\begin{equation}\label{eq:gentimeevol}
\begin{split}
        \ket{\psi(t)}&=\int_{-\infty}^\infty \diff x [\psi_L(x,t)C_L^\dag(x)\\ &+\psi_R(x,t)C_R^\dag(x)] \ket{0} + \sum_Q e_Q(t) \ket{e_Q},
\end{split}
\end{equation}
where $\psi_{R/L}(x,t)$ are right/left-moving field amplitudes, $e_Q(t)$ are the qubit excitation coefficients and $\ket{e_Q}=\sigma^\dag_Q \ket{0}$. Here, the most general single-excitation state is a superposition of a photon moving to the right and the qubits are all in the ground state ($\int_{-\infty}^\infty \diff x [\psi_R(x,t)C_R^\dag(x) \ket{0}$), a photon moving to the left and the qubits are all in the ground state ($\int_{-\infty}^\infty \diff x [\psi_L(x,t)C_L^\dag(x) \ket{0}$), the field is in vacuum state and individual qubits are in the excited state ($e_Q(t) \ket{e_Q}$). The photon wave-packet can either be generated from the qubits, or it could also come from the initial conditions as an incident pulse, or can include contributions from both cases.

The time evolution of the state $\ket{\psi(t)}$ follows the Schr\"odinger equation $i\partial_t \ket{\psi(t)}=H \ket{\psi(t)}$. After straightforward algebra (See Appendix \ref{sec:appA}), we find the time evolution equation for $\psi_{R/L}(x,t)$ and $e_Q(t)$ as
\begin{subequations} \label{eq:eom}
\begin{align}
    i (\partial_t-\partial_x) \psi_L(x,t)&=\sqrt{\frac{\gamma_0}{2}}\sum_Q e_Q(t) \delta_Q, \label{eq:eom1} \\
    i(\partial_t+\partial_x)\psi_R(x,t)&=\sqrt{\frac{\gamma_0}{2}}\sum_Q e_Q(t) \delta_Q, \label{eq:eom2} \\
    \forall Q, \, (i\partial_t-\Omega) e_Q(t) &= \sqrt{\frac{\gamma_0}{2}} \psi(x_Q,t) \label{eq:eom3}.
\end{align}
\end{subequations}
Here, we define $\psi(x,t)=\psi_R(x,t)+\psi_L(x,t)$ and $\delta_Q = \delta(x-x_Q)$ as short-hand notations. We emphasize that Eqs. (\ref{eq:gentimeevol}-\ref{eq:eom}) apply to a general case and do not contain any assumptions regarding where the qubits are located, or even how many qubits there are. With slight adjustments (to $\Omega$ and $\gamma_0$), they can also describe non-identical qubits. Using Eq. (\ref{eq:eom}), we re-derive the time-evolution equations for the single qubit \cite{fang2018non} and $N$ qubits \cite{liao2015single} from a real-space perspective in Appendices \ref{sec:app2} and \ref{sec:app3} respectively, where we correct some typos from both papers.

\begin{figure}
    \centering
    \includegraphics[width=\columnwidth]{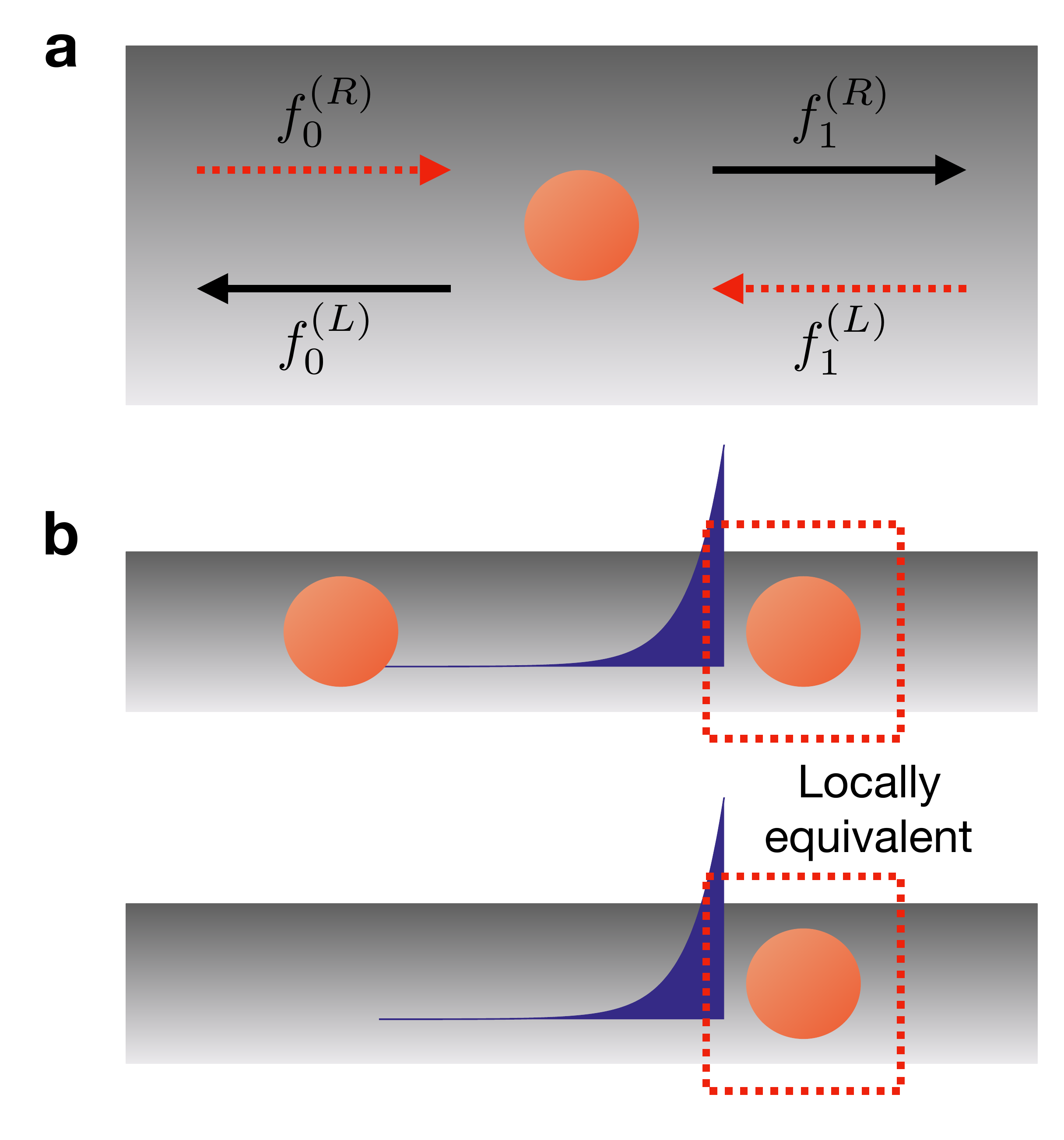}
    \caption{\textbf{a} The time evolution of pulse scattering for a single qubit. The red dashed dotted arrows correspond to incident pulses, whereas solid black arrows correspond to transmitted pulses. \textbf{b} A pulse is generated by an initially excited qubit and travels to an adjacent qubit (top), a pulse is incident from far left on a single qubit (bottom). Regardless of incident pulse's source of origin, the two scenarios are equivalent in terms of the local equations of motion at the second qubit.}
    \label{fig:fig1}
\end{figure}

\subsection{Motivating the diagrams: Locality}

Before mathematically discussing locality, we make three important remarks here:
\begin{enumerate}
    \item For $x\neq x_Q$, the field behaves like a free-field, where the field is simply translated in a shape-preserving way.
    \item The existence of source terms leads to non-continuities at the right/left moving field at the atomic position such that
    \begin{equation}
        \psi_{L/R}(x_{Q^+},t)=\psi_{L/R}(x_{Q^-},t) \pm i\sqrt{J_0} e_Q(t).
    \end{equation}
    This also suggests that while the right/left moving components are discontinuous, the field itself $\psi(x_Q,t)=\psi_L(x_Q,t)+\psi_R(x_Q,t)$ is continuous at the qubit position.
    \item The discontinuity of the right/left moving components at a qubit position depends only on the excitation of the corresponding qubit, that is, the interaction Hamiltonian derives changes in the field \emph{locally}. Similarly, the feedback from the field to the qubit is also local due to Eq. (\ref{eq:eom3}). Hence, for the scattering of a pulse from a qubit, only the shape of the incoming field \emph{at the atomic position} is important and other parameters such as the qubit number are irrelevant. The source of the incoming pulse is irrelevant, as a geometry with only a single qubit would lead to the same local EoM as the one with many qubits, which we shortly discuss.
\end{enumerate}

To illustrate the emergence of locality in equations of motion, we first make the piece-wise substitution:
\begin{subequations} \label{eq:localeqs}
\begin{align}
    \psi_R(x,t)&=\sum_Q f_Q^{(R)}([x-x_Q]-t) \\ 
    &\times [\Theta(x-x_Q)-\Theta(x-x_{Q+1})], \nonumber \\
    \psi_L(x,t)&=\sum_Q f_Q^{(L)}([x-x_Q]+t) \\
    &\times[\Theta(x-x_Q)-\Theta(x-x_{Q+1})]. \nonumber 
\end{align}
\end{subequations}
Here, $\Theta(x)$ is the Heaviside function. To cover the boundaries, one could assume that $x_{0}=-\infty$ and $x_{N+1}=\infty$. Otherwise, $x_{Q+1}=x_Q+L$. Setting these equations into Eq. (\ref{eq:eom}), we obtain
\begin{subequations} \label{eq:localeqs2}
\begin{align}
    f_Q^{(R)}(-t)-f_{Q-1}^{(R)}(L-t) &= - i \sqrt{\frac{\gamma_0}{2}} e_Q(t), \\
    f_Q^{(L)}(t)-f_{Q-1}^{(L)}(t+L) &= i \sqrt{\frac{\gamma_0}{2}} e_Q(t), \\
    (i\partial_t - \Omega)e_Q(t) &= \sqrt{\frac{\gamma_0}{2}} [f_Q^{(R)}(-t)+ f_Q^{(L)}(t)]. 
\end{align}
\end{subequations}
These are indeed local equations, which only depend on the qubit $Q$ and the incoming and out-going fields. In these equations, we can consider the incoming fields ($f_{Q-1}^{(R)}([x-x_{Q-1}]-t)$ and $f_Q^{(L)}([x-x_Q]+t)$) as incident pulses to the qubit $Q$, after which $e_Q(t)$ as well as $f_Q^{(R)}([x-x_Q]-t)$ and $f_{Q-1}^{(L)}([x-x_{Q-1}]+t)$ can be uniquely determined. We illustrate this for a single qubit in Fig. \ref{fig:fig1}a. 

In other words, we can consider the time evolution of the field-qubit system as a local phenomenon, where we consider the incoming fields as initial conditions. Of course, $f_{Q-1}^{(R)}([x-x_{Q-1}]-t)$ and $f_Q^{(L)}([x-x_Q]+t)$ do not always come from initial conditions, rather they result from the scattering of light from adjacent qubits, e.g. they are out-going pulses for some other qubit $Q'$. On the other hand, the equations of motion do not distinguish between the two cases. The field and the qubit excitation coefficients update locally, the source of the incoming field does not matter. We illustrate this in Fig. \ref{fig:fig1}b, where the two scenarios are mathematically equivalent. 

In the end, we will have developed a recursive method that starts from the initial conditions and updates the field and excitation coefficients one by one at qubit positions, by taking the output of one qubit as the input of the next. We will summarize the complicated calculations in this method in terms of simple diagrams and the corresponding set of rules. But first, we need to take a step back and recall the time evolution for a single qubit, which will be required when constructing the diagram rules.

\subsection{Single qubit time evolution}
We have considered the time evolution for a single qubit analytically in \cite{dinc2019exact}, which we summarize in this section. Let us consider a qubit situated at the position $x=0$. The time evolution of any single-excitation state can be given by \cite{dinc2019exact}
\begin{equation}
\begin{split}
        \ket{\psi(t)}= & \,{} U(t)\ket{\psi(0)}, \quad \text{where} \\
        &U(t)=\frac{1}{2\pi} \int_{-\infty}^\infty \diff k \ket{E_k}\bra{E_k} e^{-iE_kt}.
\end{split}
\end{equation}
The energy eigenstates (with energy $E_k=|k|$ and momentum $k$, for $k>0$) for this system, which describe the scattering of plane-waves from the qubit \cite{shen2005coherent}, can be given as
\begin{equation}
\begin{split}
        \ket{E_k}&=\int \diff x [e^{ikx}\Theta(-x)+t_ke^{ikx} \Theta(x)]C_R^\dag(x) \ket{0} \\
        &+ \int \diff x r_ke^{-ikx}\Theta(-x) C_L^\dag(x) \ket 0 + e_k \sigma^\dag \ket{0}.
\end{split}
\end{equation}
For $k<0$, we simply take the projection of $\ket{E_k}$ w.r.t. the origin \cite{dinc2019exact}. Here, $t_k$,$r_k$ and $e_k$ are:
\begin{subequations}
\begin{align}
    t_k &= \frac{\Delta_k}{\Delta_k+iJ_0},\\
    r_k&= \frac{-iJ_0}{\Delta_k+iJ_0},\\
    e_k&= \frac{\sqrt{J_0}}{\Delta_k+iJ_0},
\end{align}
\end{subequations}
where $\Delta_k=E_k-\Omega$ is the detuning energy. Then, the time evolution of an arbitrary state $\ket{\psi(t)}$ can be given as
\begin{equation}
    \ket{\psi(t)} = \frac{1}{2\pi} \int \diff k \braket{E_k}{\psi(t=0)}\ket{E_k} e^{-iE_k t}.
\end{equation}
This time-evolution equation will be employed in Appendix \ref{sec:app4} when deriving the diagram rules which we introduce shortly.

\begin{figure*}
    \centering
    \includegraphics[width=\textwidth]{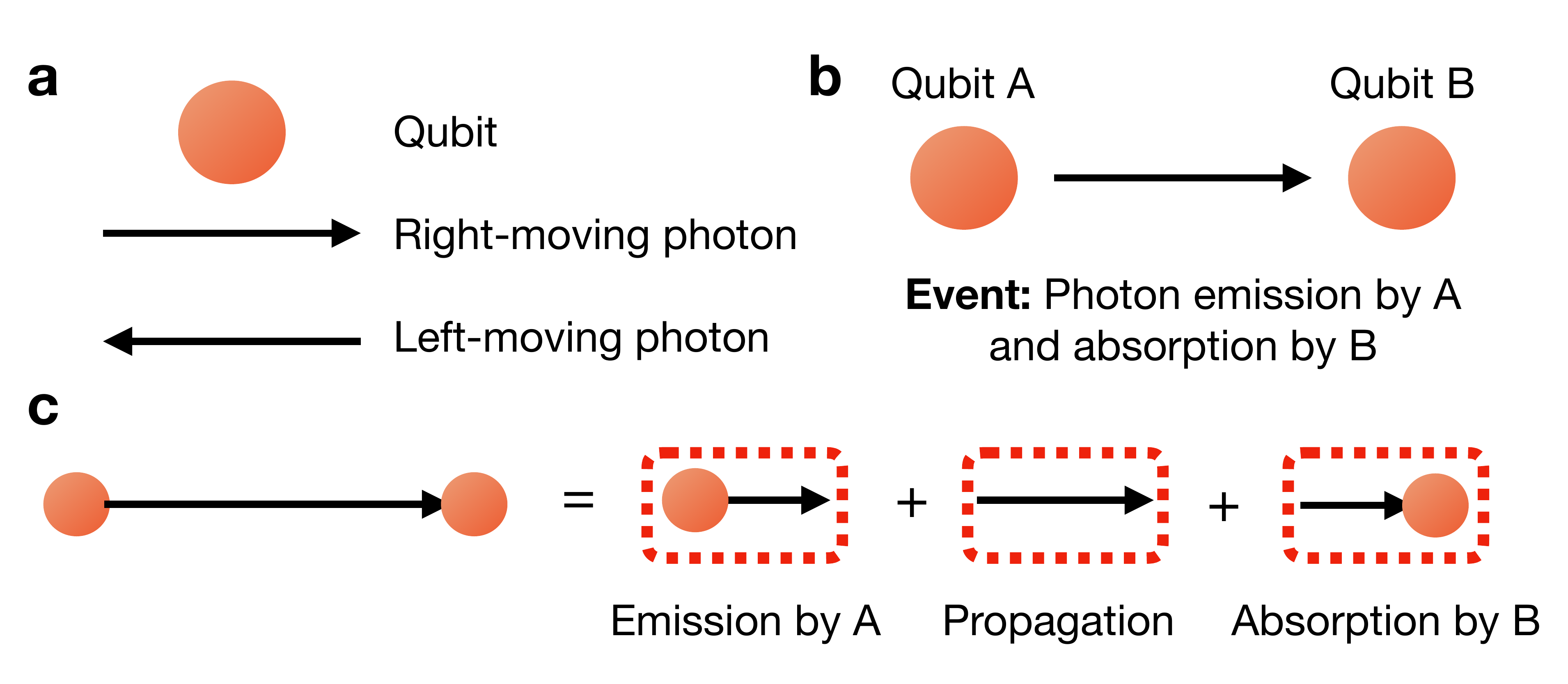}
    \caption{A heuristic illustration of diagrams. \textbf{a} There are three physical components of the diagrams: qubits and right/left-moving photons. \textbf{b} An example event whose amplitude can be calculated using the diagrams. A photon is emitted by the qubit A, propagates for some distance and gets re-absorbed by the qubit B. \textbf{c} The illustration of the event in terms of diagrams. The diagrams break down the complete event into smaller unit cells, for each cell there is a special rule to calculate the input-output relation. Each unit cell is governed by the local equations of motion, whereas cascading defines the initial conditions for the next unit cell such that the output of one unit cell is the input of another. Diagrams are convenient ways of solving Schr\"odinger equation for the single-photon subspace without any explicit computation.}
    \label{fig:fig2}
\end{figure*}

\subsection{An intuitive introduction to diagrams}
So far, we have set up the local equations of motions, established a local equivalence between multi-qubit systems and a single qubit and recalled the real-space method discussed in \cite{dinc2019exact} that allows finding the fully analytical time-evolved state for a single qubit. Our main goal is to bring these concepts together in a simple way such that we can calculate the analytical time evolution of any state in single-photon waveguide QED with multiple emitters. The missing link is provided by the diagrams that we introduce in this section, which dissect a multi-qubit system into smaller pieces that contain at most a single qubit. Since we know the time-evolution of single qubits from the previous section and Appendix \ref{sec:app4}, we can find the time-evolution of these smaller pieces and hence of the whole diagram.

We start with some simple definitions. Events are physical processes that happen in super-positions of each other, whereas diagrams, by definition, are illustrative descriptions of events that help calculating physical amplitudes. In Fig. \ref{fig:fig2}, we describe an example event including two qubits. Fig. \ref{fig:fig2}a shows the physical components of the event, which are qubits and photons. The qubits are represented with orange colored circles, whereas photons are represented with arrows. In Fig. \ref{fig:fig2}b, we describe a specific physical event where qubit A emits a photon, which is later absorbed by a second qubit B. This event is in super-position with many others that describe the spontaneous emission in the case of two qubits. We pick this one specifically, as it is the simplest event possible, hence corresponds to the most elementary diagram. Fig. \ref{fig:fig2}c shows the corresponding diagram, which contributes to the calculation of the excitation coefficient for the qubit B. 

The diagrams consist of three or more unit cells, where a unit cell is the smallest portion of the diagram that includes at most a single-qubit and for which an input-output rule exists. There are three types of unit cells: i) starters ii) propagators iii) finishers. In Fig. \ref{fig:fig2}c, there are one of each. The emission of the photon by the qubit A is a starter, propagation without interaction is a propagator, and the absorption of the photon by the qubit B is a finisher. In an arbitrary diagram, there is one starter and finisher each, but there is no limit on the number of propagators.

Before discussing the specific rules, we emphasize two important discussion points:
\begin{enumerate}
    \item Our diagrams are not Feynman diagrams, we do not consider asymptotic or steady-states, we consider the exact time evolution of an initial state. The asymptotic solutions for waveguide QED can be found using the input-output approach or many others, which have already been discussed in the literature \cite{fan2010input}. The input-output relations described by diagrams in this work are not related to this type of input-output formalism at all.
    \item The diagrams solve the time-dependent Schr\"odinger equation, whose solution is unique once the initial and boundary conditions are well-defined. Thus, the diagrams uniquely determine the exact solution.
\end{enumerate}

\begin{figure*}
    \centering
    \includegraphics[width=\textwidth]{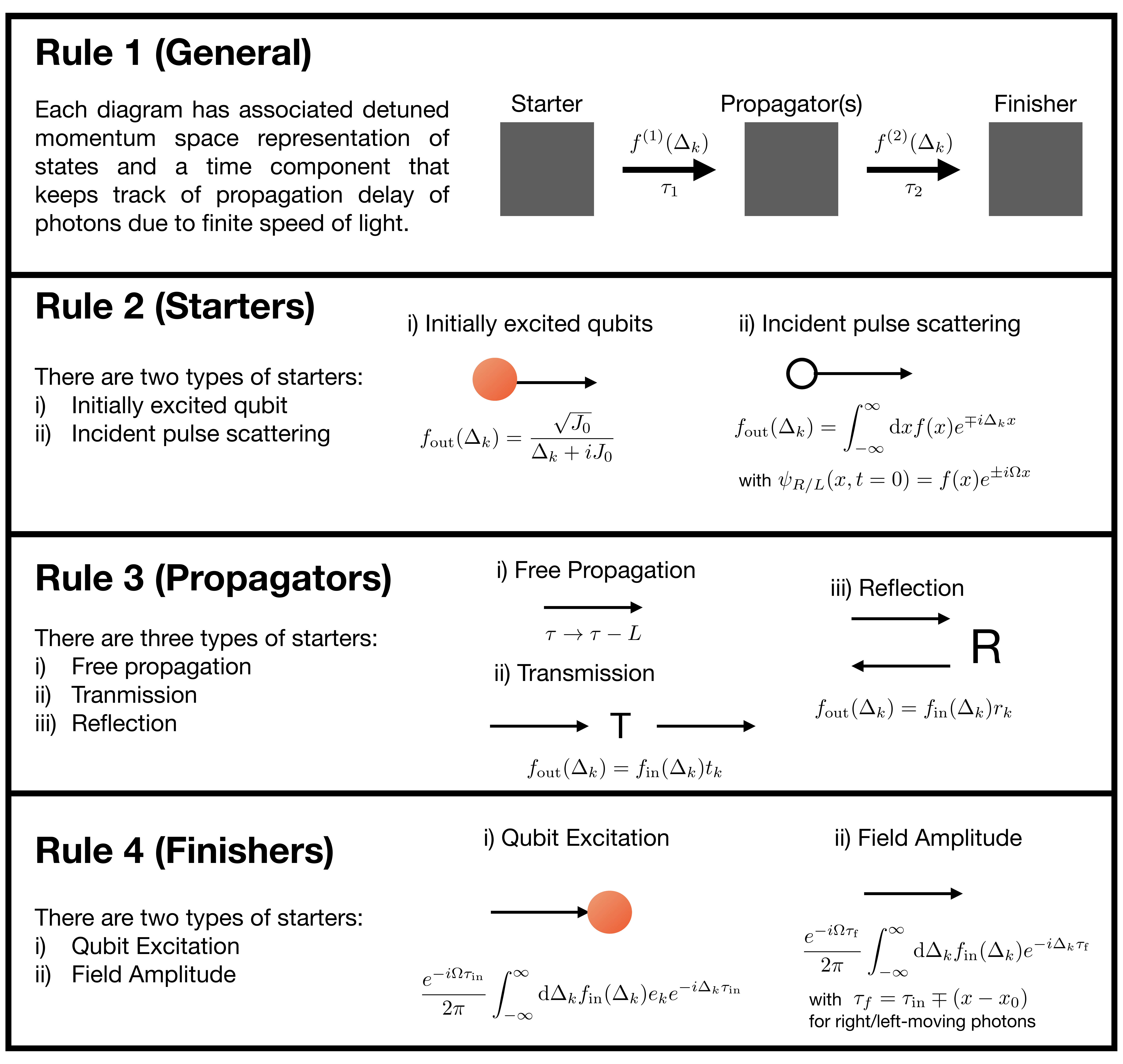}
    \caption{A visual summary for the set of diagrammatic rules. If followed, they provide exact and analytical non-Markovian time evolution for single-photon waveguide QED.}
    \label{fig:fig3}
\end{figure*}

\subsection{The set of diagrammatic rules}
In this section, we introduce the diagrammatic rules post-hoc and derive them from starting principles in Appendix \ref{sec:app4}. There is a total of four basic rules:

\begin{rules}[The unit cell compartmentalisation]
For each diagram associated with are the detuned (absolute) momentum space ($\Delta_k$) representation of states and a time component ($\tau$) that keeps track of propagation delay of photons due to finite speed of light. The diagrams are divided into smaller unit cells that keep track of both quantities. There are three types of unit cells: starters, propagators and finishers.

Momentum space component: All unit cells except for starters have an input, $f_{\rm in}(\Delta_k)$, and all except for finishers have an output, $f_{\rm out}(\Delta_k)$. Starters are associated with a unique $f_{\rm out}(\Delta_k)$, propagators have an input-output relationship between $f_{\rm in}(\Delta_k)$ and $f_{\rm out}(\Delta_k)$, finishers are associated with a unique amplitude calculation rule that determines the excitation coefficients ($e_Q(t)$) or the field-amplitudes ($\psi_{R/L}(x,t)$) utilizing $f_{\rm in}(\Delta_k)$. 

In the case of cascaded unit cells, the output of the previous cell is the input of the next:
\begin{equation}
     f_{\rm in}^{(i+1)}(\Delta_k) = f_{\rm out}^{(i)}(\Delta_k).
\end{equation}
Time component: Some propagators update the time component ($\tau$) by introducing a time-delay, the time-delay is propagated within the cascades.
\end{rules}

\begin{rules}[Starters]
There are two possible starter unit cells: i) an initially excited qubit ii) an initial pulse. The output of the starter unit cells can be given
\begin{enumerate}
    \item For an initially excited qubit:
    \begin{equation}
        f_{\rm out} (\Delta_k)=\frac{\sqrt{J_0}}{(\Delta_k+iJ_0)}.
    \end{equation}
    \item For an initial pulse with $\psi_{R/L}(x,t=0)=f(x)e^{\pm i \Omega x}$: 
    \begin{equation}
        f_{\rm out}(\Delta_k)=\int_{-\infty}^\infty \diff x f(x) e^{\mp i \Delta_k x}.
    \end{equation}
\end{enumerate}
For both starters, the time component starts as $\tau=t$.
\end{rules}

\begin{rules}[Propagators]
There are three types of propagators: i) propagation without interaction ii) transmission through a qubit iii) reflection from a qubit. The input-output relationships for the propagator unit cells can be given
\begin{itemize}
    \item For propagation without interaction for a distance $L$:
    \begin{equation}
        f_{\rm out}(\Delta_k) = f_{\rm in}(\Delta_k), \quad \tau_{\rm out} = \tau_{\rm in}-L.
    \end{equation}
    \item For transmission through a qubit:
    \begin{equation}
         f_{\rm out}(\Delta_k) = f_{\rm in}(\Delta_k) t_k, \quad \tau_{\rm out} = \tau_{\rm in}.
    \end{equation}
    \item For reflection from a qubit:
    \begin{equation}
         f_{\rm out}(\Delta_k) = f_{\rm in}(\Delta_k) r_k, \quad \tau_{\rm out} = \tau_{\rm in}.
    \end{equation}
\end{itemize}
\end{rules}

\begin{rules}[Finishers]
There are two types of finishers: i) qubit excitation coefficient ii) right/left-moving field amplitudes. The calculation of observables for the finishers can be given
\begin{enumerate}
    \item For qubit excitation coefficients
    \begin{equation}
        e_Q(t) = \frac{e^{-i\Omega \tau_{\rm in}}}{2\pi} \int_{-\infty}^\infty \diff \Delta_k f_{\rm in}(\Delta_k) e_k e^{-i\Delta_k \tau_{\rm in}}.
    \end{equation}
    \item For right/left-moving field amplitudes
    \begin{equation}
        f^{(R/L)}_Q(x,t)=\frac{e^{-i\Omega \tau_{\rm f} }}{2\pi} \int_{-\infty}^\infty \diff \Delta_k f_{\rm in}(\Delta_k) e^{-i\Delta_k \tau_{\rm f}},        
    \end{equation}
    where $\tau_{\rm f}=\tau_{\rm in} \mp (x-x_0)$ for right/left-moving field amplitudes with $x_0$ being the position of the final qubit that the photon has scattered from.
\end{enumerate}
\end{rules}

\begin{figure*}
    \centering
    \includegraphics[width=\textwidth]{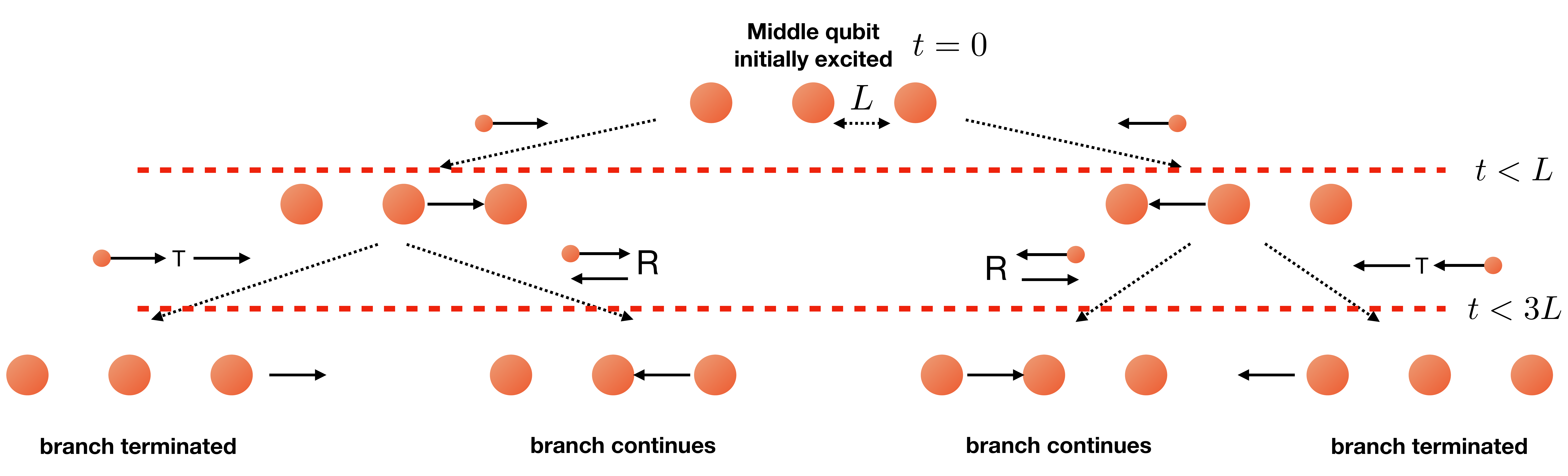}
    \caption{Systematic approach for finding the event diagrams that contribute to the amplitude calculations. For illustration purposes, we consider $N=3$ qubits with the initial condition that the middle qubit is excited. The qubits are separated by $L$. To find the event diagrams, we start with a fractal tree and branch out at every scattering point. If the photon leaves the system, the branch is terminated, as is the case for the two out of four branches at the final step. Earlier branches describe earlier times, which are noted with horizontal dotted red lines. The event diagrams are illustrated on top of the branch arrows.}
    \label{fig:fig4}
\end{figure*}

We visually summarize the diagram rules in Fig.~\ref{fig:fig3}. The diagrams give the qubit excitations $e_Q(t)$ and local field amplitudes $f_Q^{(R/L)}$, from which we can find $\psi_{R/L}(x,t)$ via Eq. (\ref{eq:localeqs}). 

\subsection{An example diagram calculation for the event in Fig. \ref{fig:fig2}b}
Using the diagrammatic rules, let us calculate the excitation coefficient for the qubit $B$ for the event given in Fig. \ref{fig:fig2}b, with the initial condition that the qubit $A$ is initially excited. We will perform the calculations step by step for illustration purposes. 

As we mentioned before, there are three unit cells in this example. Let us start with the starter:
\begin{equation}
    f_{\rm out}^{(1)}(\Delta_k)= \frac{\sqrt{J_0}}{\Delta_k+iJ_0}, \quad \tau=t.
\end{equation}
Now, by the cascading rule, we know that $f_{\rm in}^{(2)}(\Delta_k)=f_{\rm out}^{(1)}(\Delta_k)$ and by the propagator rule 
\begin{equation}
    f_{\rm out}^{(2)}(\Delta_k)=f_{\rm in}^{(2)}(\Delta_k)=\frac{\sqrt{J_0}}{\Delta_k+iJ_0}, \quad \tau=t-L. 
\end{equation}
Once again, by the cascading rule,  $f_{\rm in}^{(3)}(\Delta_k)=f_{\rm out}^{(2)}(\Delta_k)$. Finally, using the finisher rule, we obtain
\begin{equation}
\begin{split}
        e_B(t)&= \frac{e^{-i\Omega (t-L)}}{2 \pi} \int_{-\infty}^\infty \diff \Delta_k f_{\rm in}^{(3)}(\Delta_k) e_k e^{-i\Delta_k (t-L)}, \\
        &= \frac{e^{-i\Omega (t-L)}}{2 \pi} \int_{-\infty}^\infty \diff \Delta_k \frac{J_0}{(\Delta_k+iJ_0)^2} e^{-i\Delta_k (t-L)}, \\
        &= -J_0 (t-L) e^{-(J_0+i\Omega)(t-L)} \Theta(t-L).
\end{split}
\end{equation}
We have calculated the amplitude corresponding to this event. However, this specific event describes this problem only for $t<3L$. For later times, there will be other events, hence diagrams, that contribute to this amplitude.
\begin{figure*}
    \centering
    \includegraphics[width=\textwidth]{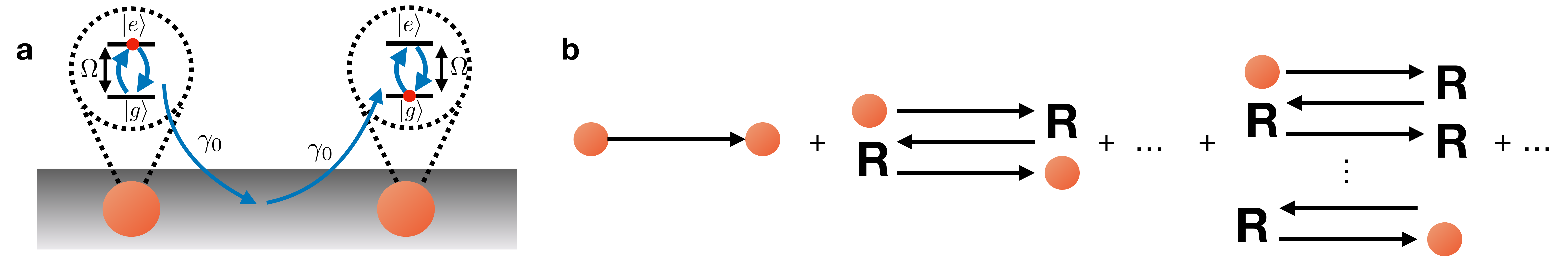}
    \caption{Fermi's two atom problem in waveguide QED \textbf{a} Two identical qubits are coupled to a one-dimensional waveguide. The first qubit is initially excited, the second one is initially in the ground state. The first qubit spontaneously decays with a rate $\gamma_0$, whereas the emitted photon is picked up by the second qubit after some time delay due to photon propagation. \textbf{b} All diagrams contributing to the qubit excitation coefficient $\braket{e_{1}}{\psi(t)}$, where $\ket{\psi(t=0)}=\ket{e_{-1}}$. Using the diagrams, one can answer Fermi's two atom problem for waveguide QED.}
    \label{fig:fig5}
\end{figure*}

\subsection{Fractal trees for finding all events}
In general, for given initial conditions and qubit configuration, there are many events that contribute to the amplitudes. The exact time-evolution for the amplitudes can be found by summing over all possible diagrams/events. This leads to a solution in terms of a time-series expansion, which is exact. Luckily, the more complicated events with multiple propagators start contributing at later times due to the propagation delay introduced within the system. Thus, the early terms in the time-series expansion can be found with simple diagrams. In fact, for a given final time $t_f$, we use a binary fractal tree to find all possible diagrams that contribute to the time-series expansion such that the exact time evolution is obtained for $t<t_f$. This is possible due to the binary nature of scattering (transmission or reflection) of light from a qubit in a one-dimensional waveguide.

We illustrate the binary fractal tree for an example of $N=3$ qubits in Fig. \ref{fig:fig4}. In this specific tree, we keep track of the right/left-moving field amplitudes, whereas the event diagrams for the excitation coefficients can be written down in a similar way. For any given final time of interest $t=t_f$, one needs to keep branching out the tree such that all the events with delays less than $t_f$ are obtained. 

Since the fractal tree branches out rapidly, it is not the most ideal way to find all events for a multi-emitter system. The diagrammatic method requires a more efficient event-finding schema to better describe systems with many emitters. This is not our focus in this paper and we leave it as a future research direction to the literature. For the rest of the paper, we turn our focus to illustrating the power of diagrams for the Fermi's two atom problem, which can be exactly and analytically solved, for the first time, in waveguide QED.

\section{Exact causality in single-photon waveguide QED}
So far, we introduced the diagrammatic approach for finding the analytical non-Markovian time evolution of states in the Schr\"odinger picture. Now, we will leverage this to answer the general question, whether exact causality is preserved in single-photon waveguide QED. In this section, we first start with Fermi's two atom problem, then generalize to any waveguide QED system. Finally, we discuss the Markov approximation that leads to short-distance causality violations inside the multi-qubit system and the consequent emergence of collective decay rates.

\subsection{Fermi's two atom problem}
Fermi's two atom problem has been considered in \cite{sabin2011fermi} for circuit QED using Heisenberg picture from a perturbative approach. Here we find the analytical time evolution in the Schr\"odinger picture, which we show to be exact in Appendix \ref{sec:app5}. Our results in this section agree with the findings of \cite{sabin2011fermi}.

To provide the first test of causality, we now focus our attention to the two-qubit system, where one of them (call it qubit $-1$) is initially in the excited state and emits a photon over time. The second one (call it qubit $1$) is in the ground state and can get excited by some finite probability after some time-delay. The test is whether the second qubit can get excited before $t<L$, as this would mean that a photon can travel faster than light and violate causality. For a causal theory, $e_1(t<L)=0$ should hold.

We now consider the configuration shown in Fig. \ref{fig:fig5}a, where the distance between qubits is $L$ and we pick the middle point of the qubits as the center of the $x$-axis. As an initial condition, we assume that the qubit $-1$ is excited $\ket{\psi(t=0)}=\ket{e_{-1}}$. Then, our aim is to find the amplitude
\begin{equation}
    e_1(t) = \braket{e_1}{\psi(t)}.
\end{equation}

The diagrams contributing to this amplitude are illustrated in Fig. \ref{fig:fig5}b, which echo those found from the perturbative expansion of observables in the Heisenberg picture \cite{sabin2011fermi}. The events corresponding to $n$th diagrams include, an emission by qubit $-1$, an absorption by qubit $1$ and $2n$ reflections with $(2n+1)$ free propagation over length $L$. As $n$ increases, we consider events, where the photon is trapped between the qubits for longer times ($(2n+1)L$ to be precise). These are lower probability events as some portion of the excitation escapes in the shape of a radiating field for every event.

Following the rules we introduced earlier and the diagrams in Fig. \ref{fig:fig5}b, we write the excitation coefficient as
\begin{equation}
    e_1(t) = \sum_{n=0}^\infty e_1^{(n)}(t),
\end{equation}
where $e_1^{(n)}(t)$ corresponds to the $n$th diagram. Then, following the diagram rules, we find
\begin{equation}
\begin{split}
    e_1^{(n)}(t) &= \frac{e^{-i\Omega \tau_n}}{2\pi} \int_{-\infty}^\infty \diff \Delta_k \frac{r_k^{2n} e_k\sqrt{J_0}}{\Delta_k +i J_0}  e^{-i\Delta_k \tau_n}, \\
    &=- \frac{(\tau_n J_0)^{2n+1}}{(2n+1)!} e^{-(J_0+i\Omega)\tau_n}\Theta(\tau_n),
\end{split}
\end{equation}
where $\tau_n = t-(2n+1)L$. Here, we performed contour integration for two cases $t > \tau_n$ and $t< \tau_n$. Bringing all together, we find
\begin{equation} \label{eq:firstqubitexcitation}
    e_1(t) = - \sum_{n=0}^\infty \frac{(\tau_n J_0)^{2n+1}}{(2n+1)!} e^{-(J_0+i\Omega)\tau_n}\Theta(\tau_n).
\end{equation}
We discuss the complete time-evolved state $\ket{\psi(t)}$ in Appendix \ref{sec:app5}, where we show that $\ket{\psi(t)}$ is indeed an exact solution of the Schr\"odinger equation, hence $e_1(t)$ found here is exact. Considering the times $t<L$, we find
\begin{equation}
    e_1(t) = 0, \quad t\leq L
\end{equation}
such that the causality is not violated for the two atom problem. Waveguide QED passes its first test.

\subsection{Generalization of causality: no-UHP theorem}
The second test of waveguide QED is a general one. What if there were more than one waveguide and arbitrary numbers of quantum emitters? Luckily, to prove causality for a general system, we do not need to introduce any more diagrams than we already have discussed. 

Consider the first diagram in Fig. \ref{fig:fig5}b, which is the earliest contribution to the excitation coefficient. Indeed, this diagram corresponds to the event we described in Fig. \ref{fig:fig2}. As we discussed then, this diagram (or an equivalent one) is the most elementary diagram which can be built. For any excitation problem, this diagram sets the lower bound on the time it takes to excite a distant qubit. Hence, regardless of the number of qubits or waveguides, the distant qubit cannot be excited before the propagation time has passed. This ensures us that for an initial condition of type 1 (when a qubit is initially excited), single-photon waveguide QED is causal.

When it comes to initial condition type 2, we do not even need the diagrams. From Schr\"odinger equation, it is clear that $H_I \ket{\psi(t)}=0$ unless the field at qubit positions is nonzero. The free Hamiltonian propagates the field with a constant speed of light, which is taken as $v_g=1$ and modulates the phase of the excitation coefficient but cannot change excitation probability. In other words, a qubit in the ground state can only be excited if the field amplitude at the qubit position is non-zero. Thus, the initial condition type 2 does not violate causality either. 

Now that we have shown that there are no causality violations in single-photon waveguide QED, we can now motivate the no-UHP theorem. Once again, we start discussing a linear chain of $N$ qubits. Later, we will argue how the results are more general than this specific case. 

Consider a linear chain of $N$ qubits, where qubits are located at $x=mL$ with $m=0,\ldots, N-1$. As an initial condition, we pick the state with an incoming decaying exponential
\begin{equation}
    \ket{\psi(t=0)}= \int_{-\infty}^{-x_0} \diff x \sqrt{2\sigma} e^{\sigma (x+x_0)} e^{i\Omega x} C_R^\dag(x) \ket{0},
\end{equation}
where $\sigma \sim O(\gamma_0)$ and $x_0>0$. Then, the time evolution of this state is proportional to \cite{dinc2019exact}
\begin{equation}
    \ket{\psi(t)}\sim \int_{-\infty}^\infty \diff \Delta_k \frac{e^{-i\Delta_k (t-x_0)}}{\Delta_k + i \sigma} \ket{E_{k^+}}.
\end{equation}
Here, we will not keep track of the proportionality factors, as they are not important for our consideration. Moreover, $\ket{E_{k^+}}$ ($k^+$ means that $k>0$) includes the scattering coefficients $e_k^{(m)}$, $r_k$ and $t_k$, as well as occasional $e^{\pm ikx}$ terms. Now, let us compute the excitation coefficient for the qubit $m$ at time $t$
\begin{equation} \label{eq:final}
    \braket{e_m}{\psi(t)} \sim \int_{-\infty}^\infty \diff \Delta_k \frac{e^{-i\Delta_k (t-x_0)}}{\Delta_k + i \sigma} e_k^{(m)}.
\end{equation}
As $x_0 \to \infty$, the field is far away from the qubits. The field amplitude at any qubit position is zero, hence $\braket{e_m}{\psi(t)}=0$. On the other hand, as $x_0 \to \infty$, the contour of the integral in Eq. (\ref{eq:final}) is connected through the upper half-plane. For the integral to be zero, regardless of the time $t$, the coefficient $e_k^{(m)}$ should not have any upper half-plane poles. Similar arguments can be made for $t_k$ and $r_k$ if we consider the field amplitudes instead of the excitation coefficients. 

For a more general system with many waveguides, the same argument can be made by picking one waveguide and an initial pulse such as the decaying exponential as in Eq. (\ref{eq:final}). The reason behind choosing the decaying exponential is the following: The decaying exponential is zero for $x>-x_0$, hence cannot excite qubits until later times $t>x_0$. Its momentum representation has only one lower half-plane pole at $\Delta_k = - i \sigma$ and no upper-half plane poles. Then, following the discussion in \cite{dinc2019exact} regarding a general waveguide QED system, integral given in Eq. (\ref{eq:final}) describes the amplitudes with the only exception that $e_k^{(m)}$ is replaced with the corresponding scattering parameter, generalizing our findings regarding causality in linear chains.

Now, we are ready to state the no-UHP theorem:
\begin{theorem}[No-UHP Theorem]
For any waveguide QED system with multiple emitters and waveguides, the scattering parameters do not have any upper half-plane poles. 
\end{theorem}
Thus, for any scattering matrix calculation, we expect the resulting matrix elements to have poles only on the lower half-plane or on the real axis. This is a waveguide QED analogy to previously known causality theorems in quantum theory \cite{gellmann1954use,schutzer1951onthe}.

\subsection{Time-delayed coherent quantum feedback, short-distance causality violations and emergent collective decay rates}
\begin{figure}
    \centering
    \includegraphics[width=\columnwidth]{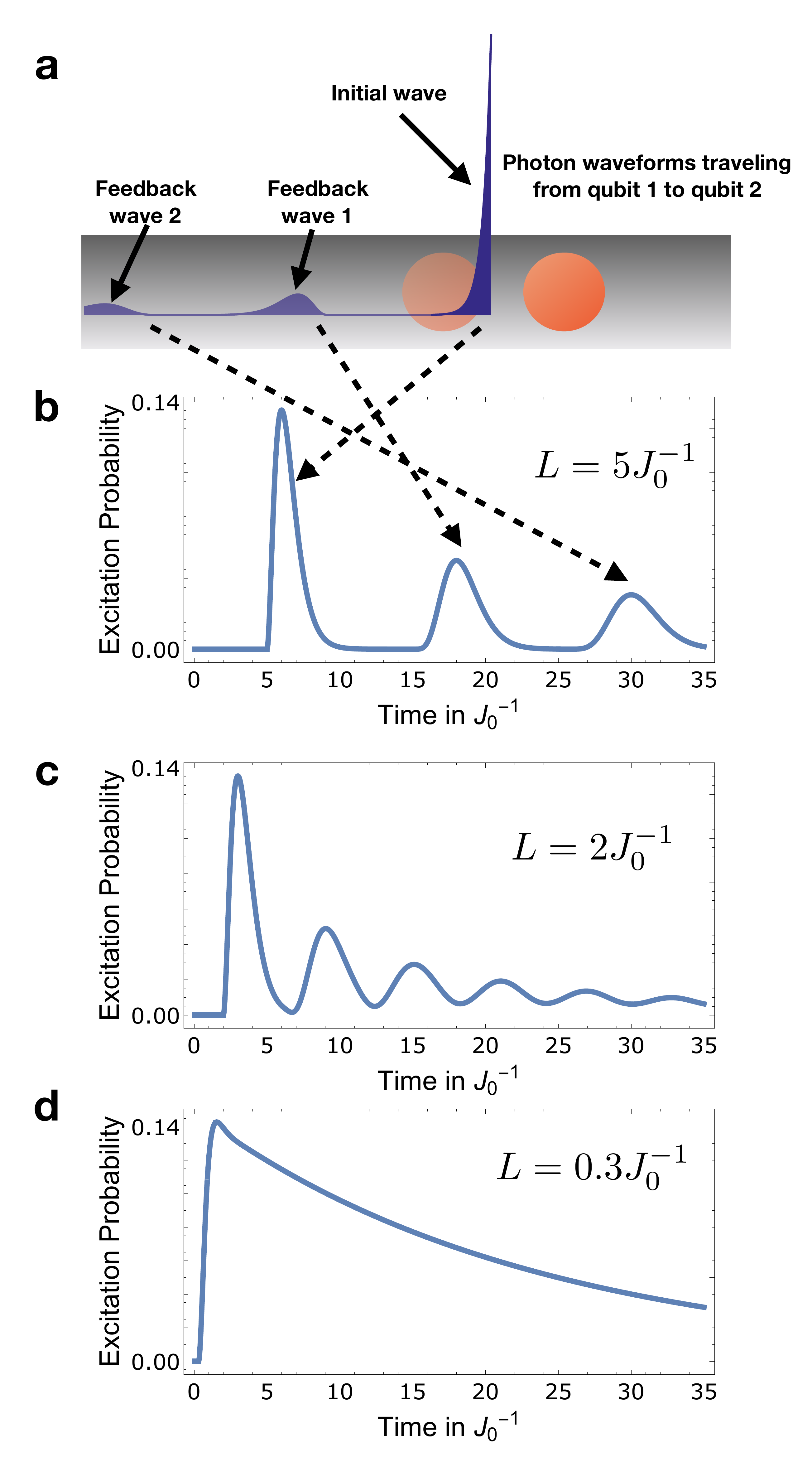}
    \caption{The time-delayed feedback leads to the periodic re-excitation of the second qubit with time intervals of $2L$. \textbf{a} The field mediates the interaction between the two qubits with some time delay. Here, only the field ongoing from qubit 1 to the qubit 2 is shown, where the pale blue corresponds to the future state of the trapped photonic field moving to the right. The feedback waves arise from the internal reflections of the photon. The field amplitudes are illustrated using analytical expressions calculated in Appendix \ref{sec:app5}. \textbf{b} The second qubit gets excited periodically due to the incoming feedback waves for a large separation of $L=5J_0^{-1}$. \textbf{c} The qubits are moderately separated with $L=2J_0^{-1}$, where interference between different feedback waves can be observed. \textbf{d} The qubits are closely separated with $L=0.3J_0^{-1}$ such that from the interference between nearly-instantaneous feedback waves emerges a collective behavior. For all plots, we pick $\Omega = 200 J_0$.}
    \label{fig:fig6}
\end{figure}

Having discussed causality for single-photon waveguide QED, we now turn our attention to a commonly used Markov approximation, where short-distance causality violations are allowed for algebraic convenience when solving the time-evolution of the model. Within this approximation, the time-evolution of single-photon states can be found fully analytically \cite{dinc2019exact}, making it a powerful tool for considering microscopically separated qubits where the quantum coherent feedback mediated by the field is nearly-instantaneous. In this section, we first visualize the feedback for largely separated, even isolated, qubits. Then, using the analytical expressions from Appendix \ref{sec:app5}, we show the emergence of collective decay rates for the two-qubit system in the Markovian limit $L \to 0$ (or more practically, $L \to O(\Omega^{-1})$ with $J_0 \ll \Omega$). In this limit, we ignore any causality violations within $O(\Omega^{-1})$, as any events with such short time duration are assumed to happen instantaneously. 

Now, let us consider the interactions in waveguide QED systems from a qualitative point of view. An excited qubit decays to the waveguide, which takes the form of the field excitation. The field propagates with a finite speed $v_g$ (which is taken $v_g=1$ for convenience) and scatters inside the multi-qubit system. A portion of the excitation travels back to the initial qubit with a certain time-delay introduced by the finite propagation speed, re-excites the qubit and the feedback process goes on infinitely many times. We call the portion of the field that re-excites the initial qubit \emph{the feedback waves} (See Fig. \ref{fig:fig6}a). In many cases, the feedback waves get dampened over time such that less and less excitation remains inside the system and more gets radiated away.\footnote{In a very special case, the dampening does not happen, and a finite amount of excitation gets trapped inside the system signaling the existence of so-called \emph{bound-states in continuum} for the specific waveguide QED system \cite{calajo2019exciting}} In this highly qualitative picture, the field trapped between qubits mediates the interactions in a time-delayed manner; and has a memory built-into it through all the scattering inside the system and through the finite, non-zero width of the photon wave-packet. This is, partially, the reason why we primarily keep track of the field scattering in the event diagrams (both in Figures \ref{fig:fig4} and \ref{fig:fig5}) even when we are calculating the qubit excitation coefficients. 

Depending on how early or late the consecutive feedback waves arrive back to the initial qubit, the qubits can show a various range of collective behavior. If there is a sufficiently long time (much longer than $J_0^{-1}$) between two feedback waves (See Fig. \ref{fig:fig6}b), the qubits are nearly isolated and not much collective behavior is observed. As the qubit separation decreases (See Figs. \ref{fig:fig6}c-d), the subsequent feedback wave travels back to the original qubit before the decay of the previous wave, leading to time-retarded constructive or destructive interference between multiple consecutive feedback waves. The qubits can no longer be viewed as single isolated entities, but rather they are in waveguide-mediated interactions with each other. These interactions lead to novel collective phenomena such as superradiance, subradiance and bound-states in continuum \cite{dinc2019exact}. 

In the limit, where the feedback happens almost instantly, there is almost no field stored between the qubits. This is because $L\sim(\Omega^{-1})$ and $|\psi(x,t)|^2  \sim O(J_0) $ (from diagram rules and Appendix \ref{sec:app5}) such that the probability amplitude of the trapped photon is vanishingly small $\mathcal{P}(x,x+L) \approx L |\psi(x,t)|^2  \sim O(J_0/\Omega) \ll 1$. Thus, the trapped field no longer stores the \emph{memory} of the feedback waves, as they almost instantly arrive. Consequently, we call this limit the \emph{Markovian regime} \cite{dinc2019exact}. 

Now, let us re-consider the two-qubit example, specifically Eq. (\ref{eq:firstqubitexcitation}) and show that we can obtain the collective decay rates from this exact expression when we take the Markovian limit. We take the limit $L \to O(\Omega^{-1})$ such that $J_0 L \to 0$, $t \sim O(J_0^{-1})$ and define the phase parameter $\theta = \Omega L$. Then, we obtain for $t>0$
\begin{equation}
\begin{split}
        &e_1(t) =  - \sum_{n=0}^\infty \frac{(\tau_n J_0)^{2n+1}}{(2n+1)!} e^{-(J_0+i\Omega)\tau_n}\Theta(\tau_n), \\
        &\approx  -e^{-(J_0+i\Omega)t} \sum_{n=0}^\infty \frac{(t J_0)^{2n+1}}{(2n+1)!} e^{i\theta(2n+1)} \Theta(t),\\
        &=-e^{-(J_0+i\Omega)t} \sum_{n=0}^\infty \frac{(t J_0e^{i\theta})^{2n+1}}{(2n+1)!}, \\
        &= -\frac{e^{-i\Omega t}}{2} e^{-J_0 t} \sinh(t J_0 e^{i\theta}),\\
        &= \frac{e^{-i\Omega t}}{2} \left( e^{-\frac{\Gamma_1}{2}t} -e^{-\frac{\Gamma_2}{2}t} \right),
\end{split}
\end{equation}
where we read-off the \emph{collective decay rates} as
\begin{subequations}
\begin{align}
    \Gamma_{1/2} = \gamma_0 (1\pm e^{i\theta}).
\end{align}
\end{subequations}
These are indeed the collective decay rates of the two-qubit system \cite{dinc2019exact}. We note, however, in this limit, the second qubit gets excited almost simultaneously after the decay of the first qubit. This leads us to conclude:
\begin{theorem}
Causality is violated for short-distances in the Markovian limit, which can describe the microscopically separated qubits well.
\end{theorem}
Thus, the price of obtaining an analytically convenient and physically intuitive collective picture made available through the Markov approximation is the acceptance of short-distance violations in the system time-evolution.

\section{Conclusion}

Non-Markovian time evolution in waveguide QED is a challenging problem that needs to be tackled in order to aid the theoretical research in waveguide QED. So far, the literature has employed various assumptions and/or numerical approaches to tackle this problem. Among those assumptions, Markovianity has been the most prominent. If qubits are microscopically separated, the propagation time delay between qubits can be neglected (Markovianity assumption). This regime is fairly well-explored in the literature \cite{dinc2019exact} and provides important insights on the collective behavior of the system. If the qubits are separated by large distances, then each qubit behaves in an isolated manner, hence collective effects become unimportant. This regime is easily explorable, as, in this regime, the decay of one qubit happens before another one has the chance to be excited. Interestingly, the mysteries lie at the moderate separation, where both time-delayed feedback and collective interactions dictate the behavior of the system. In fact, interesting phenomena had been observed when multiple qubits are separated moderately, including but not limited to enhanced quantum memory \cite{calajo2019exciting} or enhanced superradiance (or so-called super-superradiance) \cite{dinc2019non,sinha2020non}. 

In this work, we have tackled the non-Markovian time evolution problem in single-photon waveguide QED. We have developed a diagrammatic approach, which breaks down the complicated computations into a simple set of rules governing unit cells. With this method, we consider the waveguide QED equivalent of Fermi's two atom problem, show that causality is not violated in single-photon waveguide QED, and  conclude with a discussion on Markov approximation commonly employed in the waveguide QED literature and consequent short-distance causality violations introduced to the time-evolution. In our considerations of causality, we also provide the literature with the no-UHP theorem, which states that S-matrix calculations should not lead to scattering parameters with upper half-plane poles. Both the diagrams and the no-UHP theorem can be beneficial as a sanity check for theoretical research in waveguide QED.

Further research on diagrams is necessary when we consider large systems with many qubits, as the number of possible diagrams increases rapidly following a fractal tree (See, for example, Fig. \ref{fig:fig4}). As future work, a more efficient process is required to find all the diagrams that contribute to the calculations of amplitudes. On the other hand, in such large systems, the early time dynamics can still be exactly captured by the early stages of the fractal tree. Therefore, the diagrams found from the tree can aid the theoretical research in waveguide QED especially in bench-marking theoretical models, as any theory should reproduce the exact results obtained from them. Hence, we provide the literature, for the first time, with the analytical ground truth of the non-Markovian time evolution of single-excitation states driven by the point-coupling Hamiltonian in waveguide QED. 

\section*{Acknowledgements}
The author would like to thank the Perimeter Scholars International Essay Defense Committee, Eduardo Martinez, Tibra Ali and Francois David, for insightful questions regarding the causality in real-space formalism; which eventually lead to this work, and Agata Bra\'nczyk and Lauren E. Hayward for supervising the essay.

\onecolumngrid
\appendix
\section{Derivation of Eq. (\ref{eq:eom})}\label{sec:appA}
We first note that the state $\ket{\psi(t)}$ in Eq. (\ref{eq:gentimeevol}) is a solution to the Schr\"odinger equation
\begin{equation}
    i \partial_t \ket{\psi(t)}=H \ket{\psi(t)}.
\end{equation}
Left-hand-side can be found as
\begin{equation}
  i\partial_t \ket{\psi(t)}=  i\int_{-\infty}^\infty \diff x [\partial_t \psi_L(x,t) C_L^\dag(x)+\partial_t \psi_R(x,t)C_R^\dag(x)] \ket{0} +i \sum_Q \partial_t e_Q(t) \ket{e_Q}.
\end{equation}
Then, let us find the action of the free Hamiltonian
\begin{equation}
    H_0 \ket{\psi(t)} = i\int_{-\infty}^\infty \diff x [\partial_x \psi_L(x,t) C_L^\dag(x)-\partial_x \psi_R(x,t)C_R^\dag(x)] \ket{0} + \Omega \sum_Q e_Q(t) \ket{e_Q}.
\end{equation}
Before moving forward with the interaction Hamiltonian, let us assume that $\gamma_0=0$ and consider the case when there is no interactions. In this case, the EoM are as follows
\begin{subequations} \label{eq:eomwithoutI}
\begin{align}
    \partial_t \psi_L(x,t)&=\partial_x \psi_L(x,t), \\
    \partial_t \psi_R(x,t)&=-\partial_x \psi_R(x,t), \\
    \partial_t e_Q(t) &= -i\Omega e_Q(t).
\end{align}
\end{subequations}
The general solution to these equations are as follows
\begin{equation}
    \psi_{L/R}(x,t)=f(t \pm x), \quad e_Q(t)=e_Q(0) e^{-i\Omega t}, 
\end{equation}
where $e_Q(0)$ are the initial conditions and $f$ is any pulse shape that is either left moving (for L) or right moving (for R). In fact, for L, this represents a pulse moving to left with a constant speed while keeping its shape constant, whereas for R, it is a solitonic pulse moving to right. In short, the free Hamiltonian acts as a translator for the photonic component and as a phase modulator for the qubit components. 

Let us now consider the interaction Hamiltonian, which gives
\begin{equation*}
    H_I \ket{\psi(t)}=\sqrt{\gamma_0/2} \sum_Q (\psi_L(x_Q,t)+\psi_R(x_Q,t)) \ket{e_Q} + \sqrt{\gamma_0/2} \sum_Q e_Q(t) C_L^\dag(x_Q) \ket{0} + \sqrt{\gamma_0/2} \sum_Q e_Q(t) C_R^\dag(x_Q)\ket{0}.
\end{equation*}
Adding the interaction terms to Eq. (\ref{eq:eomwithoutI}), we obtain Eq. (\ref{eq:eom}).

\section{Time-evolution for a single qubit and causality} \label{sec:app2}
Fang et al \cite{fang2018non} consider the non-Markovian dynamics of a single qubit coupled to a one-dimensional waveguide. In doing so, they also derive the time evolution equation for the qubit excitation coefficient in the single-excitation sector in Appendix B.1. In this Appendix, we fix the typo performed in this Appendix, provide the accurate solution and correct their incorrect claim on causality.

We shall start by re-stating the arguments of \cite{fang2018non}. First, the authors consider the time evolution of an arbitrary state in the single-excitation sector and find the following coupled differential equations
\begin{subequations} \label{eq:singlequbit}
\begin{align} 
    i \partial_t \phi_R(x,t) &= -i \partial_x \phi_R(x,t) + Ve(t)\delta(x), \\
    i \partial_t \phi_L(x,t) &= i \partial_x \phi_L(x,t) + Ve(t)\delta(x), \\
    i \frac{\diff}{\diff t} e(t) &=\Omega e(t) +V [\phi_R(0,t)+\phi_L(0,t)]. 
\end{align}
\end{subequations}
Here, $\phi_{R/L}(x)$ are right/left-moving field amplitudes, $e(t)$ is the excitation coefficient of the qubit, $\Omega$ is the excited state energy, $V$ is the coupling between the qubit and the continuum ($\sqrt{\gamma_0/2}$ for us) and $\delta(x)$ is the Dirac-delta. Then, the authors claim to solve the equations (\ref{eq:singlequbit}a-b) to obtain the following:
\begin{subequations} \label{eq:singlequbit2}
\begin{align}
    \phi_R(x,t)&=\phi_R(x-t,0)-iVe(t-x)\Theta(x) \Theta(t-x),  \\
    \phi_L(x,t)&=\phi_L(x+t,0)-iVe(t+x)\Theta(-x) \Theta(t+x).
\end{align}
\end{subequations}
Here, the authors claim the causality is preserved, which is expected thanks to $\Theta(t\pm x)$. However, these theta functions are indeed typos. A quick check would be to check whether (\ref{eq:singlequbit2}a) solves (\ref{eq:singlequbit}a):
\begin{equation}
    i (\partial_t + \partial_x) \phi_R(x) = V e(t) \delta(x) \Theta(t) \neq V e(t) \delta(x).
\end{equation}
Similar expression can be obtained for Eq. (\ref{eq:singlequbit2}b). Now, as long as we consider times $t>0$, most arguments followed by \cite{fang2018non} are consistent since $\Theta(t)=1$ for $t>0$. This is precisely the reason why the authors do not encounter any major issues, as they consider time evolution for $t>0$. On the other hand, their claim on the immediate manifestation of causality preservation by $\Theta(t \pm x)$ is not accurate and therefore their interpretation of causality is incorrect. We will comment on this aspect later.

Then, the authors set Eqs. (B3) into Eq. (B2c) to obtain
\begin{equation} \label{eq:singlequbit3}
    \frac{\diff}{\diff t} e(t) =- \left[i\Omega+ \frac{\gamma_0}{2} \right]e(t)-iV [\phi_R(-t,0)+\phi_L(t,0)]. 
\end{equation}
Here, $\gamma_0=2V^2$ is the decay rate of a single emitter. On one hand, it is common to consider initial value problems in waveguide QED. The issue with Fang et. al's derivation does not lie with the assumption $t>0$, and indeed their end result for the time evolution equation is correct for an initial value problem. The issue is with the interpretation provided by the typo, which is incorrect even if the problem is posed as an initial value problem. On another hand, we believe that this is a good opportunity to derive the full time evolution equation and therefore we will provide the complete derivation. 

Let us first show that the time evolution equation in \cite{fang2018non} leads to non-physical results. Consider Eq. (\ref{eq:singlequbit3})  under the condition that $\phi_R(-t,0)=\phi_L(t,0)=0$ and $e(0)=1$. Then, let us re-define $\alpha(t)=e(t)e^{-i\Omega t}$, leading to the equation:
\begin{equation}
    \frac{\diff}{\diff t} \alpha(t) = \frac{\gamma_0}{2} \alpha(t).
\end{equation}
Under the picked initial conditions, we find $\alpha(t)= \exp(-\frac{\gamma_0 t}{2})$. However, this solution is not physical, since $\alpha(t)$ blows up as $t \to -\infty$. In fact, the expected solution would be \cite{rephaeli2010full}
\begin{equation} \label{eq:singlequbit4}
    \alpha(t) =\exp(-\frac{\gamma_0}{2}|t|)= \exp(-\frac{\gamma_0}{2}t)\Theta(t)+\exp(\frac{\gamma_0}{2}t)\Theta(-t).
\end{equation}
We shall show that this is indeed the solution to the equation derived without the typo.

Now, we fix the typo by re-writing Eq. (\ref{eq:singlequbit2}) in the following correct form:
\begin{subequations} \label{eq:singlequbit6}
\begin{align}
    \phi_R(x,t)&=\phi_R(x-t,0)+iVe(t-x) \Theta(x-t) - iVe(t-x)\Theta(x), \\
    \phi_R(x,t)&=\phi_L(x+t,0)+iVe(t+x) \Theta(-x-t) - iVe(t+x)\Theta(-x).
\end{align}
\end{subequations}
Now, we find the summation $\phi_R(0,t)+\phi_L(0,t)=\phi_R(x-t,0)+\phi_L(x+t,0)-iVe(t) \sign(t)$, where we define $\sign(x)=1-2\Theta(-x)=\Theta(x)-\Theta(-x)$ and $\Theta(0)=0.5$ as pointed out by \cite{fang2018non}. Then, we find the time evolution equation for the excitation coefficient as
\begin{equation}
    \frac{\diff}{\diff t} e(t) = - \left( i \Omega + \frac{\gamma_0}{2}\sign(t)\right)e(t)-iV [\phi_R(-t,0)+\phi_L(t,0)].
\end{equation}
The difference between this equation and the Eq. (B4)  is the existence of $\sign(t)$ function. Now, again, let us set $e(t) = \alpha(t) e^{-i\Omega t}$, $\phi_R(-t,0)=\phi_L(t,0)=0$ and $e(0)=1$ to obtain:
\begin{equation} \label{eq:singlequbit5}
    \frac{\diff}{\diff t} \alpha(t) = - \frac{\gamma_0}{2} \sign(t) \alpha(t).
\end{equation}
To solve this equation, we can first assume $t>0$, which would give $\alpha(t)=\exp(-\frac{\gamma_0}{2}t)$ and then assume $t<0$, which would give $\alpha(t)=\exp(\frac{\gamma_0}{2}t)$. Combining both, we find the solution from Eq. (\ref{eq:singlequbit4}). Indeed, $\alpha(t)$ does not blow up as $t \to \pm \infty$, as expected from a physical system. It is straightforward to show that Eq. (\ref{eq:singlequbit4}) is a solution to Eq. (\ref{eq:singlequbit5}).

Now that we have fixed the typo in Eq. (\ref{eq:singlequbit2}), this might raise a serious question. Would the time evolution be no longer causal, as the incorrect terms in Eq. (\ref{eq:singlequbit2}) are said to be "preserving the causality?" The causality for a single-qubit system has been considered in \cite{dinc2019exact} and indeed is preserved. This can also be seen from our Eq. (\ref{eq:singlequbit6}), which is the corrected version of Eq. (\ref{eq:singlequbit2}). To consider $t>0$, we assume that the excitation coefficient has the form $e(t)=\xi(t)\Theta(t)$. The theta function in this assumed form provides the theta functions dropped from  Eq. (\ref{eq:singlequbit2}). 

So far, we have focused on the single-excitation sector. However, a similar issue stands for the double-excitation sector considered in Appendix B.2 of \cite{fang2018non}. We will not discuss this further due to brevity reasons, one can perform a quick check by setting Eq. (B8a) of \cite{fang2018non} into (B7c) of \cite{fang2018non}. Once again, this does not change the main conclusions of the paper under the assumption $t>0$, only the interpretation of the source of causality.

\section{Derivation of generalized time-evolution equation of excitation coefficients for a linear chain of qubits} \label{sec:app3}
Liao et al \cite{liao2015single} consider the time evolution of a linear chain of qubits coupled to a one-dimensional waveguide. The authors derive the time evolution equation for the qubit excitation coefficients upon a single-photon pulse scattering. However, there is a mistake in their derivations of time evolution equation due to careless handling of integral boundaries. In this Appendix, we identify the mistake and then correct it. First, we intuitively point to the problem in their time evolution equation. Second, we derive the correct time evolution equation from a real space perspective and, third, we comment on where the mistake happens in \cite{liao2015single}. In summary, the missing term in Eq. (8) of \cite{liao2015single} becomes important when non-Markovian effects (from time-delayed coherent quantum feedback) become important.

Let us start by stating the time evolution equation derived by Liao. et. al., which is Eq. (8) of \cite{liao2015single}:
\begin{equation} \label{eq:M8}
    \dot \alpha_j(t)=b_j(t)-\sum_{n=0}^{N-1} \frac{\gamma_0}{2} e^{i\Omega |n-j| L } \alpha_n\left(t- |n-j|L\right),
\end{equation}
where $\alpha_j(t)$ is the (adjusted with $e^{-i\Omega t}$) excitation coefficient of qubit $j$ at time $t$, $b_j(t)$ is a contribution coming from the initial incident pulse, $\gamma_0$ is the single emitter decay rate, $L$ is the separation between two qubits, $\Omega$ is the excited state energy for the qubits. We assume no non-radiative decay, as it is irrelevant to our discussion. 

After we define the initial conditions for $t=0$, we realize from Eq. (\ref{eq:M8}) that they do not uniquely determine the time evolution. One needs to know the $\alpha_j(t)$ values even for some time $t<0$. Consider the following case: We are interested in time interval $t\in [0,\epsilon]$, where $\epsilon$ is a very small number. Then, the RHS of Eq. (\ref{eq:M8}) would require that we know the excitation probability $\alpha_j(-O(L))$, where $O(L)$ means at the order of $L$. This is alarming, because we are used to setting up Schr\"odinger equation as an initial value problem and Eq. (\ref{eq:M8}) requires a continuum of initial values for $-(N-1)L<t<0$ on top of the values at $t=0$. \cite{liao2015single} does not discuss this issue, but since the qubits are initially at the ground state and the pulse is initially far away with effectively zero field amplitude at qubit positions, it is not far off to assume a ground state initial condition for the time interval before $t=0$. In any case, Eq. (\ref{eq:M8}) raises enough suspicion to motivate diving into the derivations.

In this Appendix, we will derive the time evolution equation from a real-space perspective. The derivation performed by \cite{liao2015single} can still be re-traced to obtain the correct time evolution, which we will do in the end of this Appendix. For now, we will introduce another perspective to the problem.

We start by finding the solution to Eq. (\ref{eq:eom1}) and (\ref{eq:eom2}) as:
\begin{subequations}
\begin{align}
        \psi_{L}(x,t)&=\psi_L(x+t,0)+i \sqrt{\frac{\gamma_0}{2}} \sum_Q e_Q(t+[x-x_Q]) \Theta(-[x-x_Q]-t) -i \sqrt{\frac{\gamma_0}{2}} \sum_Q e_Q(t+[x-x_Q]) \Theta(-[x-x_Q]), \\
        \psi_R(x,t)&= \psi_R(x-t,0) +i \sqrt{\frac{\gamma_0}{2}} \sum_Q e_Q(t-[x-x_Q])\Theta([x-x_Q]-t )- i \sqrt{\frac{\gamma_0}{2}} \sum_Q e_Q(t-[x-x_Q])\Theta(x-x_Q),
\end{align}
\end{subequations}
where we define $\Theta(0)=0.5$. One can check this solution post-hoc by setting these into Eq. (\ref{eq:eom}).

Now, we can set this into Eq. (\ref{eq:eom3}) to obtain
\begin{equation} \label{eq:qubitQproof}
\begin{split}
      (i \partial_t - \Omega) e_Q(t) &= \sqrt{\frac{\gamma_0}{2}} \Big[\psi_R(x_Q-t,0)+\psi_L(x_Q+t,0)\Big] \\
      &- i \frac{\gamma_0}{2} \left[ \sum_{x_{Q'}<x_Q} e_{Q'}(t-[x_Q-x_{Q'}]) +\sum_{x_{Q'}>x_Q} e_{Q'}(t+[x_Q-x_{Q'}]) + e_{Q}(t) \right]  ,\\
      &+i\frac{\gamma_0}{2} \sum_{Q'} \left[ e_{Q'}(t+[x_{Q}-x_{Q'}])\Theta(-[x_Q-x_{Q'}]-t)+e_{Q'}(t-[x_Q-x_{Q'}])\Theta([x_Q-x_{Q'}]-t )  \right], \\
      &= \sqrt{\frac{\gamma_0}{2}} \Big[\psi_R(x_Q-t,0)+\psi_L(x_Q+t,0)\Big] - i \frac{\gamma_0}{2}  \sum_{n=0}^{N-1} e_n(t-|n-Q|L)\\
      &+i\frac{\gamma_0}{2} \sum_{n=0}^{N-1} [ e_n (t + (Q-n)L)\Theta(-(Q-n)L-t)+ e_n(t-[Q-n]L)\Theta((Q-n)L-t) ].
\end{split}
\end{equation}
We can further simplify the last term in the above expression:
\begin{equation} 
\begin{split}
      &\sum_{n=0}^{N-1} [ e_n (t + (Q-n)L)\Theta(-(Q-n)L-t)+ e_n(t-[Q-n]L)\Theta([Q-n]L-t) ] =, \\
      &= \left[ \sum_{n=Q}^{N-1} e_n(t+ [Q-n]L) \Theta(- [Q-n]L-t ) + \sum_{n=0}^{Q-1} e_n(t-[Q-n]L) \Theta([Q-n]L-t)   \right] \\
      &+ \left[ \sum_{n=0}^{Q-1} e_n(t+ [Q-n]L) \Theta(- [Q-n]L-t ) + \sum_{n=Q}^{N-1} e_n(t-[Q-n]L) \Theta([Q-n]L-t)   \right], \\
      &= \sum_{n=0}^{N-1} e_n(t-|n-Q|L)\Theta(|n-Q|L-t) + \sum_{n=0}^{N-1} e_n(t+|n-Q|L)\Theta(-|n-Q|L-t).
\end{split}
\end{equation}
We can now bring all together to obtain the general time-evolution equation:
\begin{equation}
\begin{split}
        i (\partial_t - \Omega)e_Q(t)&=\sqrt{\frac{\gamma_0}{2}} \Big[\psi_R(x_Q-t,0)+\psi_L(x_Q+t,0)\Big] - i \frac{\gamma_0}{2}  \sum_{n=0}^{N-1} e_n(t-|n-Q|L) \Theta(t-|n-Q|L) \\
        &+i \frac{\gamma_0}{2} \sum_{n=0}^{N-1} e_n(t+|n-Q|L) \Theta(-|n-Q|L-t).
\end{split}
\end{equation}
Now, this equation is the complete time evolution equation for all times $t$. \cite{liao2015single} considers $t>0$, so we shall apply this condition, which makes the final term zero, use the initial conditions $\psi_R(x,0)=\phi_R(x)$ and $\psi_L(x,0)=0$ and define $e_j(t)=\alpha_j(t) e^{-i\Omega t}$ to obtain
\begin{equation}
    \dot \alpha_j(t) = -i\sqrt{\frac{\gamma_0}{2}} \phi_R(x_j-t) e^{i\Omega t} - \frac{\gamma_0}{2} \sum_{n=0}^{N-1} \alpha_n(t-|n-j|L) e^{i\theta|n-j|}\Theta(t-|n-j|L).
\end{equation}
where $\dot \alpha(t)=\partial_t \alpha(t)$, $\theta=\Omega L$. This is the correct form of Eq. (8) of \cite{liao2015single} derived from the real space perspective in the absence of non-radiative decay. The only difference is the existence of $\Theta(t-|n-j|L)$ functions multiplied with $\alpha_n(t-|n-j|L)$. The corrected equations do not require a continuum of initial conditions, only those defined at $t=0$. Both equations agree for $t>(N-1)L$, as the $\Theta(t-|j-n|L)$ term becomes one. However, for finite and low $t$, the two equations do not agree.

Let us now see why the derivation by \cite{liao2015single} does not agree with ours. This is because in moving from Eq. (6) to Eq. (8), \cite{liao2015single} does not carefully consider the integral boundaries. During their calculations, they encounter an integral: 
\begin{equation}
    \int_0^t \diff \tau \delta(\tau-|i-j|L)=
    \begin{cases}
    1 \quad t>|i-j|L, \\
    0 \quad t<|i-j|L.
    \end{cases}
\end{equation}
For large $t$, $|i-j|L$ falls within the integral bounds $[0,t]$ and therefore the integral is evaluated as one. However, for low $t$ values, this integral evaluates to zero. Hence, this integral brings a Heaviside function $\Theta(t-|i-j|L)$ which supplies the missing term in Eq. (\ref{eq:M8}).

Now, let us consider the Markovian regime, where the qubits are microscopically separated. If we define $J_{jl}= e^{i\theta |j-l|}$ and perform the Markovian approximation ($\alpha[t-O(L)]\simeq \alpha[t]$) and $\Theta(t-|n-j|L)\approx\Theta(t)=1$ for $t>0$, we obtain the matrix equation for the time evolution
\begin{equation}
    \dot \alpha_j(t) = -i\sqrt{\frac{\gamma_0}{2}} \phi_R(x_j-t) e^{i\Omega t}-\frac{\gamma_0}{2} \sum_l J_{jl} \alpha_l(t).
\end{equation}
The same result is obtained starting from Eq. (\ref{eq:M8}). Since the authors consider microscopic separation between qubits (which is $L \sim O(\Omega^{-1})$), their calculations fall under the Markovian regime and their main results are indeed accurate. The danger exists if one wants to use Eq. (\ref{eq:M8}) to describe non-Markovian time dynamics. In that case, the missing Heaviside function should be added. Alternatively, one could use the methods described in \cite{dinc2019exact}, where the non-Markovian time evolution boils down to computing a single numerical integral instead of solving time-delayed differential equations numerically. 

\cite{dinc2019exact} also provides a link to \cite{liao2015single} by showing that Markovian collective decay rates calculated for a linear chain from the time evolution equation agree with the poles of the scattering parameters. This discussion remains unaffected from the error corrected in this Appendix, as the link to \cite{liao2015single} is established only in the Markovian regime, which is also why the error was not spotted in \cite{dinc2019exact}.

\section{Deriving and discussing the diagram rules} \label{sec:app4}
In this Appendix, we start from the time-evolution of states for a single-qubit and derive the diagram rules along the way. 

Rule 1 is more of a descriptive rule than a mathematical statement. It picks up our discussion from Eq. (\ref{eq:localeqs2}) and provides a recipe for how to use these equations to compartmentalize the diagrams into unit cells. Here, each unit cell includes at most one single qubit, where a unit cell considers the EoM in Eq. (\ref{eq:localeqs2}) instead of Eq. (\ref{eq:eom}) upon the substitution of Eq. (\ref{eq:localeqs}). Here, an important distinction needs to be made between $\psi_{R/L}$ and $f_Q^{(R/L)}$. The former can be obtained from the latter by constraining in space $[x_Q,x_{Q+1}]$ with Heaviside functions. 

As a matter of fact, unit cell compartmentalisation is rooted in the local equivalence illustrated in Fig. \ref{fig:fig1}. For any distant qubit (distant from qubit $Q$) radiating photons, we extend the field to the complete spatial domain by considering $f_Q^{(R/L)}$ over $\psi_{R/L}$ and ignore any other qubit other than $Q$. This performs the compartmentalisation for us. The cascading part of the Rule 1 simply connects the unit cells by equating the out-going field of one cell to the incoming-field of another. This way, $f_Q^{(R/L)}$ are all connected through cascading and uniquely determined, given that there is a proper initial condition. 

The starters provide the initial condition for the local EoM, where they provide the initial $f_X^{(R/L)}$ values for some qubit $X$, other $f_Q^{(R/L)}$ become non-zero one-by-one after $f_X^{(R/L)}$ has scattered long enough. The free propagator connects the local EoMs for qubits $Q$ and $Q+1$ by introducing the time-delay it takes for the out-going field of one to be the incoming field of another. The other two propagators solve the local EoM for the qubit $Q$ by properly scattering the incoming field. The finishers either give excitation coefficients of qubits upon the incident pulse, or provide the real-space representation of the field whose detuned momentum space representation we already know. As the interactions are mediated via photons, the communication between unit cells is dependent on incoming and out-going fields. Thus, Rule 1 implicitly suggests that the diagrams follow the path of a photon.

To prove Rule 2, we start by considering the starter with an incident pulse, as it is easier to prove:
\begin{equation} \label{eq:rmphoton}
    \ket{S(t=0)}=\int \diff x f(x) e^{i\Omega x} C_R^\dag(x) \ket{0}.
\end{equation}
Here, $f(x)$ is an initially incident right-moving pulse with $f(x>0)=0$. Then, the time evolution is
\begin{equation}
\begin{split}
        \ket{S(t)} = \frac{1}{2\pi} \int_{-\infty}^\infty \diff \Delta_k \braket{E_{k^+}}{S(0)} \ket{E_{k^+}} e^{-i E_k t} , \quad \braket{E_{k^+}}{S(0)} =  \int_{-\infty}^\infty \diff x f(x) e^{-i\Delta_k x}.
\end{split}
\end{equation}
At this point, we define $f_{\rm out}(\Delta_k) = \braket{E_{k^+}}{S(0)}$, where $\ket{E_{k^+}}$ represents energy-eigenstates with positive momentum $k$. This definition will be clear when we are discussing rule 4. For now, we motivate this definition by arguing that $\braket{E_{k^+}}{S(0)}$ is the only initial condition dependent part of the integral expression, thus it is the only part that needs to be propagated within the diagram through unit cells. For a more general integral of this form, we can read-out $f_{\rm in/out}(\Delta_k)$ and $\tau$ as follows
\begin{equation} \label{eq:finaldef}
    \frac{1}{2\pi} \int_{-\infty}^\infty \diff \Delta_k  f_{\rm in/out}(\Delta_k) \ket{E_{k^+}} e^{-iE_k  \tau},
\end{equation}
which is true for right-moving photons. For left-moving photons $\ket{E_{k^+}}$ is replaced with $\ket{E_{k^-}}$.

Now, we can consider the field amplitude emitted by an initially excited qubit. For simplicity, we assume that the qubit is located at $x=0$ and we consider the emitted field for $x>0$. The initial condition for this system is 
\begin{equation}
    \ket{\psi(t=0)}=\ket{e}.
\end{equation}
Then, by the time-evolution operator in Eq. (\ref{eq:gentimeevol}), we find the time-evolved state as
\begin{equation}
    \ket{\psi(t)}= \frac{1}{2\pi} \int_{-\infty}^\infty \diff k \braket{E_k}{e} \ket{E_k} e^{-iE_kt}.
\end{equation}
Here, the integral is over the momentum $k$, not detuned energy $\Delta_k$. It is important to discuss here why we call $\Delta_k$ as the detuned (absolute) momentum in the diagrams. Since $\hbar=v_g=1$, $\Delta_k = |k| - \Omega$. Thus, $\Delta_k$ is the detuned absolute momentum as much as it is the detuned energy. We choose to pick the notion momentum over energy to put emphasis on the path of photons for the diagrams.

Now, we can find the emitted field amplitude for $x>0$ as
\begin{equation}
    S(x,t)=\braket{x}{\psi(t)} = \frac{e^{-i\Omega(t-x)}}{2\pi} \int_{-\infty}^\infty \diff \Delta_k e_k^* (t_k+r_k) e^{-i\Delta_k (t-x)} = \frac{e^{-i\Omega(t-x)}}{2\pi} \int_{-\infty}^\infty \diff \Delta_k \frac{\sqrt{J_0}}{\Delta_k+iJ_0} e^{-i\Delta_k (t-x)}.
\end{equation}
As we mentioned above, we motivate the diagrams from the path that photons follow. Thus, we prefer turning an excited qubit to an incident pulse using the local equivalence. Then, replacing the qubit with the field it radiates ($S(x,t)=\braket{x}{S(t)}$), which could become an input for a cascaded unit cell, we would have obtained for the input $\braket{E_k}{S}=\frac{\sqrt{J_0}}{\Delta_k+iJ_0}$. Therefore, we say that an initially excited qubit is equivalent to an incident pulse with
\begin{equation}
    f_{\rm out}(\Delta_k)=\frac{\sqrt{J_0}}{\Delta_k+iJ_0},
\end{equation}
with the only difference that the qubit radiates this pulse to two various direction, both of which contributes to the possible events as shown in Fig. \ref{fig:fig4}. In both cases, $\tau$ starts as $\tau=t$. 

Let us now consider the Rule 3. We start with the free propagator over the distance $x_0$, as it is the simplest one to prove. The free propagator simply shifts the reference frame by $x_0$, to the left for right-moving photons, or to the right for left moving photons. 

For now, let us discuss a right-moving photon in Eq. (\ref{eq:rmphoton}). We first let the field propagate freely for a time $t=x_0$, which gives us
\begin{equation}
    \ket{S(t=x_0)}= \int \diff x f(x-x_0) e^{i\Omega (x-x_0)} C_R^\dag(x) \ket{0}.
\end{equation}
which evolves over time as
\begin{equation}
    \ket{S(t)} = \frac{1}{2\pi} \int_{-\infty}^\infty \diff \Delta_k \braket{E'_{k^+}}{S(x_0)} \ket{E'_{k^+}} e^{-i E_k (t-x_0)}.
\end{equation}
Here, $\ket{E_k'}$ is in a different reference frame\footnote{There is no prime on $E_k$ and $\Delta_k$, as $E_k=|k|$ is independent of reference frame shifts.} than $\ket{E_k}$. We need to shift the initial reference frame by $x_0$ such that the end-point of the propagator unit cell becomes the center. This is required because when we connect the unit cells, they need to have the same reference frame and all unit cells start at the local coordinate $x'=0$. The easiest way to perform this shift is to pick $\ket{E_k}$ for a qubit located at $x=x_0$, rather than $x=0$. Then, we find
\begin{equation}
\begin{split}
        \braket{E'_{k^+}}{S(x_0)} &= \int \diff x \braket{E'_{k^+}}{x} \braket{x}{S(x_0)} = \int \diff x e^{-ik (x-x_0)} f(x-x_0) e^{i\Omega (x-x_0)}, \\
        &= \int \diff x e^{-ik x} f(x) e^{i\Omega (x)} = f_{\rm in} (\Delta_k).
\end{split}
\end{equation}
Now, the time-evolution integral becomes
\begin{equation}
    \ket{S(t)} = \frac{1}{2\pi} \int_{-\infty}^\infty \diff \Delta_k f_{\rm in}(\Delta_k) \ket{E'_{k^+}} e^{-i E_k (t-x_0)},
\end{equation}
where we can read-out $f_{\rm out}(\Delta_k) = f_{\rm in}(\Delta_k)$ and $\tau_{\rm out}=\tau_{\rm in}-x_0$, where $\tau_{\rm in}=t$ for this specific case. A similar calculation can be performed for left-moving photons, where we consider the neighbor qubit at $x=-x_0$ rather than at $x=x_0$ and $\ket{E_{k^+}}$ is replaced with $\ket{E_{k^-}}$.

Now, we can turn our attention to the other two propagator unit cells. Once again, we assume a right moving photon as in Eq. (\ref{eq:rmphoton}) that gets scattered from a qubit. The transmitted field amplitude for $x>0$ can be found as
\begin{equation}
    S_t(x,t) = \frac{1}{2\pi} \int_{-\infty}^\infty \diff \Delta_k f_{\rm in}(\Delta_k) t_k  e^{-iE_k (t-x)}.
\end{equation}
We note that $S_t(x<0,t)=0$. However, we can once again employ the local equivalence and replace the system with the qubit by the system without the qubit but the field amplitude extended to $x<0$. Then, the output field corresponds to the state:
\begin{equation}
    \ket{S_t(t)} = \frac{1}{2\pi} \int_{-\infty}^\infty \diff \Delta_k f_{\rm in}(\Delta_k) t_k \ket{E_{k^+}}  e^{-iE_k t}.
\end{equation}
We can read-out $f_{\rm out}(\Delta_k) = f_{\rm in}(\Delta_k) t_k$ and $\tau_{\rm out}=\tau_{\rm in}$, where $\tau_{\rm in}=t$ for this specific case. A similar argument can be made for the reflected pulse, where we consider $S_r(x,t)$ instead of $S_t(x,t)$ and extend it to $x>0$ with the local equivalence.

Now, we are ready to consider the final rule, Rule 4. Once again, we start with a pulse given in Eq. (\ref{eq:rmphoton}) and find the qubit excitation probability
\begin{equation}
    \braket{e}{S(t)} = \frac{1}{2\pi} \int_{-\infty}^\infty \diff \Delta_k f_{\rm in}(\Delta_k) \braket{e}{E_{k^+}} e^{-i E_k t} = \frac{e^{-i\Omega t}}{2\pi} \int_{-\infty}^\infty \diff \Delta_k f_{\rm in}(\Delta_k) e_k e^{-i \Delta_k t}.
\end{equation}
This is indeed the finisher rule for the qubit excitation coefficient, with $\tau=t$ is picked by the initial condition in Eq. (\ref{eq:rmphoton}). For a more general initial condition, $\tau$ comes from the previous parts of the diagram. The finisher rule for the field amplitudes can be found similarly by finding $\braket{x}{S(t)}$ instead of $\braket{e}{S(t)}$. 

We now see why we have propagated only $\braket{E_{k^{\pm}}}{S(t=0)}$ rather than the whole expression in the integral in Eq. (\ref{eq:finaldef}). The changes within unit cells occur only in the $f(\Delta_k)$ and $\tau$ portions of the integrals. Thus, rather than propagating the complete integral throughout the cascades, we put the integral into the finisher rules. 

\section{Time evolution in Fermi's two atom experiment} \label{sec:app5}
In this Appendix, we present the complete time evolution for the Fermi's two atom problem, which is partially discussed in the main text. We show that the diagrams give the time evolved state, which is an exact solution of the time-dependent Schr\"odinger equation.

Let us start by writing down the Hamiltonian $H=H_0 + H_I$ \cite{dinc2019exact}
\begin{subequations}
\begin{align}
    H_0 &= \Omega \sum_{j=\{-1,1\}} \ket{e_i} \bra{e_i} + i \int_{-\infty}^\infty \diff x \left( C_L^\dag(x) \frac{\partial}{\partial x} C_L(x) -C_R^\dag(x) \frac{\partial}{\partial x} C_R(x)  \right), \\
    H_I &= \sqrt{J_0}\sum_{j=\pm1} \left( \sigma_j^\dag [C_L(jL/2) + C_R(jL/2)] + \sigma_j [C_L^\dag(jL/2) + C_R^\dag(jL/2)] \right),
\end{align}
\end{subequations}
where we pick $x=0$ as the center of the $2$ qubit system. The distance between qubits is $L$. We can write down the time-evolved state as sum of many parts:
\begin{equation}
    \ket{\psi(t)}=\ket{\psi_{-1}(t)}+\ket{\psi_1(t)}+\ket{\psi_{R_i}(t)}+\ket{\psi_{R_e}(t)}+\ket{\psi_{L_i}(t)}+\ket{\psi_{L_e}(t)}. 
\end{equation}
Here, $\pm 1$ stand for qubits, $i/e$ stand for internal and external. The time evolution for these states can be found from the diagrams. We will state them here directly rather than deriving them:
\scriptsize
\begin{subequations}
\begin{align*}
    \ket{\psi_{-1}(t)}&= \sum_{n=0}^\infty \frac{(\tau_n^{(1)} J_0)^{2n}}{(2n)!} e^{-(J_0+i\Omega)\tau_n^{(1)}} \Theta(\tau_n^{(1)}) \ket{e_1}, \quad &\tau_n^{(1)} &= t-2Ln,\\
    \ket{\psi_1(t)}&=-\sum_{n=0}^\infty \frac{(\tau_n^{(2)}J_0)^{2n+1}}{(2n+1)!}e^{-(J_0+i\Omega)\tau_n^{(2)}} \Theta(\tau_n^{(2)}) \ket{e_2},\quad &\tau_n^{(2)}&=t-(2n+1)L, \\
    \ket{\psi_{R_i}(t)}&=-i\sqrt{J_0}\sum_{n=0}^\infty \int_{-\infty}^\infty \diff x \frac{(J_0 \tau_n^{(3)})^{2n}}{(2n)!} e^{-(J_0+i\Omega)\tau_n^{(3)}} \Pi_L(x) \Theta(\tau_n^{(3)}) C_R^\dag(x) \ket{0}, \quad &\tau_n^{(3)}&=(t-2nL)-(x+L/2), \\
    \ket{\psi_{R_e}(t)}&=i \sqrt{J_0} \sum_{n=0}^\infty \int_{-\infty}^\infty \diff x \frac{(\tau_n^{(4)}J_0)^{2n}}{(2n+1)!} [\tau_n^{(4)}J_0-(2n+1)]e^{-(J_0+i\Omega)\tau_n^{(4)}}\Theta(x-L/2) \Theta(\tau_n^{(4)})C_R^\dag(x)\ket{0}, \quad &\tau_n^{(4)}&=(t-[2n+1]L)-(x-L/2), \\
    \ket{\psi_{L_i}(t)}&=i \sqrt{J_0}\sum_{n=0}^\infty \diff x \frac{(\tau_n^{(5)}J_0)^{2n+1}}{(2n+1)!}e^{-(J_0+i\Omega)\tau_n^{(5)}}\Pi_L(x)\Theta(\tau_n^{(5)})C_L^\dag(x) \ket{0}, \quad &\tau_n^{(5)}&= (t-[2n+1]L)+(x-L/2), \\
    \ket{\psi_{L_e}(t)}&=-i\sqrt{J_0} \sum_{n=0}^\infty \int_{-\infty}^\infty \diff x \frac{(\tau_n^{(6)}J_0)^{2n-1}}{(2n)!} (\tau_n^{(6)}J_0-2n)e^{-(J_0+i\Omega)\tau_n^{(6)}} \Theta(-[x+L/2])\Theta(\tau_n^{(6)})C_L^\dag(x) \ket{0}, \quad & \tau_n^{(6)}&=(t-2nL)+(x+L/2).
\end{align*}
\end{subequations}
\normalsize
Here, $\Pi_L(x)=\Theta(x+L/2)-\Theta(x-L/2)$ constrains the field excitations to the internal of the two-qubit-system and the disparity between the odd and even powers of $(\tau_n^{(1/2)} J_0)$ in $\ket{\psi_{\mp 1}(t)}$ simply mean that the photon gets reflected odd or even times before being absorbed.

To prove that the EoM given in Eq. (\ref{eq:eom}) are satisfied, we first extract the amplitudes $\psi_{R/L}(x,t)$ and $e_Q(t)$ 
\begin{subequations}
\begin{align}
    \psi_L(x,t)&=i \sqrt{J_0} \sum_{n=0}^\infty \Big[ \frac{(\tau_n^{(5)}J_0)^{2n+1} }{(2n+1)!}e^{-(J_0+i\Omega)\tau_n^{(5)}} \Pi_L(x) \Theta(\tau_n^{(5)}) \\ 
    &-\frac{(\tau_n^{(6)}J_0)^{2n-1}}{(2n)!} (\tau_n^{(6)}J_0-2n)e^{-(J_0+i\Omega)\tau_n^{(6)}} \Theta(- [x+L/2])\Theta(\tau_n^{(6)}) \Big] \nonumber , \\
    \psi_R(x,t)&=i \sqrt{J_0} \sum_{n=0}^\infty \Big[ \frac{(\tau_n^{(4)} J_0)^{2n}}{(2n+1)!} [\tau_n^{(4)}J_0-(2n+1)] e^{-(J_0+i\Omega)\tau_n^{(4)}}\Theta(x-L/2) \Theta(\tau_n^{(4)})\nonumber \\
    &- \frac{(\tau_n^{(3)}J_0)^{2n}}{(2n)!}e^{-(J_0+i\Omega)\tau_n^{(3)}} \Pi_L(x) \Theta(\tau_n^{(3)}) \Big], \\
    e_{-1}(t)&=\sum_{n=0}^\infty \frac{(\tau_n^{(1)}J_0)^{2n}}{(2n)!} e^{-(J_0+i\Omega)\tau_n^{(1)}}\Theta(\tau_n^{(1)}), \\
    e_{1}(t)&=- \sum_{n=0}^\infty \frac{(\tau_n^{(2)}J_0)^{2n+1}}{(2n+1)!} e^{-(J_0+i\Omega)\tau_n^{(2)}}\Theta(\tau_n^{(2)}).
\end{align}
\end{subequations}

Then, let us check Eq. (\ref{eq:eom1})
\begin{equation}
\begin{split}
    i (\partial_t - \partial_x)\psi_L(x,t)&=\sqrt{J_0}\sum_{n=0}^\infty \Big[ \frac{(\tau_n^{(5)}J_0)^{2n+1} }{(2n+1)!}e^{-(J_0+i\Omega)\tau_n^{(5)}} [\delta(x+L/2)-\delta(x-L/2)] \Theta(\tau_n^{(5)}) \\ 
    &+\frac{(\tau_n^{(6)}J_0)^{2n-1}}{(2n)!} (\tau_n^{(6)}J_0-2n)e^{-(J_0+i\Omega)\tau_n^{(6)}} \delta( x+L/2)\Theta(\tau_n^{(6)}) \Big], \\
    &=\sqrt{J_0}\sum_{n=0}^\infty \Big[ \frac{(t-(2n+2)L)J_0)^{2n+1} }{(2n+1)!}e^{-(J_0+i\Omega)(t-(2n+2)L)} \Theta((t-(2n+2)L)) \delta(x+L/2)\\
    &- \frac{(t-(2n+1)L)J_0)^{2n+1} }{(2n+1)!}e^{-(J_0+i\Omega)(t-(2n+1)L)} \Theta((t-(2n+1)L)) \delta(x-L/2)\\ 
    &+\frac{(t-2nL)J_0)^{2n-1}}{(2n)!} ((t-2nL)J_0-2n)e^{-(J_0+i\Omega)(t-2nL)} \delta( x+L/2)\Theta(t-2nL) \Big], \\
    &= \sqrt{J_0} \delta(x+L/2) \Bigg[e^{-(J_0+i\Omega)t}\Theta(t) +\sum_{n=1}^\infty \Theta(t-2nL)e^{-(J_0+i\Omega)(t-2nL)} \Big[ \frac{((t-2nL)J_0)^{2n-1}}{(2n-1)!}\\
    &+ \frac{((t-2nL)J_0)^{2n-1}}{(2n)!} ((t-2nL)J_0-2n) \Big]\Bigg] + \sqrt{J_0}e_{1}(t) \delta(x-L/2), \\
    &= \sqrt{J_0} \delta(x+L/2) \Bigg[e^{-(J_0+i\Omega)t}\Theta(t) +\sum_{n=1}^\infty \Theta(t-2nL)e^{-(J_0+i\Omega)(t-2nL)} \frac{((t-2nL)J_0)^{2n}}{(2n)!} \Bigg] \\
    &+ \sqrt{J_0}e_{1}(t) \delta(x-L/2), \\
    &= \sqrt{J_0}e_{-1}(t) \delta(x+L/2)+\sqrt{J_0}e_{1}(t) \delta(x-L/2).
\end{split}
\end{equation}
Now, let us check Eq. (\ref{eq:eom2})
\begin{equation}
    \begin{split}
        i(\partial_t+\partial_x)\psi_R(x,t)&=- \sqrt{J_0} \sum_{n=0}^\infty \Big[ \frac{((t-(2n+1)L) J_0)^{2n}}{(2n+1)!} [(t-(2n+1)L)J_0-(2n+1)] e^{-(J_0+i\Omega)(t-(2n+1)L)}\\
    &\delta(x-L/2) \Theta(t-(2n+1)L) + \frac{((t-(2n+1)L)J_0)^{2n}}{(2n)!}e^{-(J_0+i\Omega)(t-(2n+1)L)} \delta(x-L/2) \Theta(t-(2n+1)L) \\
    &- \frac{((t-2nL)J_0)^{2n}}{(2n)!}e^{-(J_0+i\Omega)(t-2nL)} \delta(x+L/2) \Theta(t-2nL) \Big],\\
    &= \sqrt{J_0} e_1(t) \delta(x+L/2) - \sqrt{J_0}\delta(x-L/2)\sum_{n=0}^\infty e^{-(J_0+i\Omega)(t-(2n+1)L)}\Theta(t-(2n+1)L) \\
    &\frac{((t-(2n+1)L)J_0)^{2n+1}}{(2n+1)!}, \\
    &= \sqrt{J_0} e_{-1}(t) \delta(x+L/2) +\sqrt{J_0} e_{1}(t) \delta(x-L/2).
    \end{split}
\end{equation}
Now let us check Eq. (\ref{eq:eom3}) for qubit $-1$
\begin{equation}
    \begin{split}
    (i \partial_t - \Omega)e_{-1}(t)&= -i J_0 e^{-(J_0+i\Omega)t} \Theta(t)+ i e^{-(J_0+i\Omega)t}\delta(t) -i J_0 \sum_{n=1}^\infty \frac{((t-2nL)J_0)^{2n}}{(2n)!} e^{-(J_0+i\Omega)(t-2nL)} \Theta(t-2nL), \\
    &+i\sum_{n=1}^\infty \frac{((t-2nL)J_0)^{2n}}{(2n)!} e^{-(J_0+i\Omega)(t-2nL)}\delta(t-2nL) \\
    &+ iJ_0 \sum_{n=1}^\infty \frac{((t-2nL)J_0 )^{2n-1}}{(2n-1)!} e^{-(J_0+i\Omega)(t-2nL)}\Theta(t-2nL), \\
    &=iJ_0\sum_{n=0}^\infty \frac{((t-2nL)J_0)^{2n-1}}{(2n)!} [2n-(t-2nL)J_0] e^{-(J_0+i\Omega)(t-2nL)} \Theta(t-2nL) + i \delta(t), \\
    &= \sqrt{J_0}\psi_L([-L/2]^-,t) + i \delta(t).
    \end{split}
\end{equation}
Here, in the final step, we use the fact that $\psi(x,t)=\psi_R(x,t)+\psi_L(x,t)$ is continuous and is equal to $\psi_L(x_Q^{-},t)$ at $x_Q=-L/2$. Moreover, a $\delta(t)$ term occurs on top the expected terms according to Eq. (\ref{eq:eom3}). This delta term comes from the way we introduce the initial condition and EoM are satisfied with the understanding that we are considering the time evolution for $t>0$ for the initial condition $\ket{\psi(0)}=\ket{e_{-1}}$. 

Now let us check Eq. (\ref{eq:eom3}) for qubit $1$
\begin{equation}
    \begin{split}
    (i \partial_t - \Omega)e_{1}(t)&= iJ_0 \sum_{n=0}^\infty \frac{ ( [t-(2n+1)L]J_0 )^{2n+1} }{ (2n+1)! } e^{-(J_0+i\Omega) [t-(2n+1)L] }\Theta(t-(2n+1)L) \\
    &-iJ_0\sum_{n=0}^\infty \frac{ ((t-(2n+1)L)J_0 )^{2n} }{ (2n)! } e^{-(J_0+i\Omega)(t-(2n+1)L)} \Theta(t-(2n+1)L), \\
    &= i J_0 \sum_{n=0}^\infty \frac{ ( [t-(2n+1)L]J_0 )^{2n} }{ (2n+1)! }( [t-(2n+1)L]J_0-(2n+1)) e^{-(J_0+i\Omega) (t- (2n+1)L)} \Theta(t- (2n+1)L ) , \\
    &= \sqrt{J_0} \psi_R([L/2]^+,t).
    \end{split}
\end{equation}
The $\delta(t)$ function does not show up here, as the initial condition was introduced as $\ket{\psi(0)}=\ket{e_{-1}}$, which only includes the qubit on the left.  Thus, we have proven that the diagrams give the exact time-evolved states for Fermi's two atom problem.

\end{document}